\documentclass[10pt,twoside,twocolumn]{IEEEtran}
\newlength{\figwidth} 
\newlength{\caffigwidth} 

\usepackage{cite}
\usepackage[T1]{fontenc}
\usepackage{graphicx}
\usepackage{amssymb}
\usepackage{amsmath}
\usepackage{amsthm}
\usepackage{subfigure}
\usepackage{booktabs} 
\usepackage{multirow}
\usepackage{microtype}
\usepackage{balance}
\usepackage{xcolor}
\usepackage[hidelinks]{hyperref}
\usepackage{mathdots}
\usepackage{epstopdf}
\usepackage{algorithm}
\usepackage{algpseudocode}
\usepackage{setspace}
\usepackage{footmisc}
\usepackage{tikz}
\usepackage{circledsteps}
\usepackage{mdframed}
\usepackage{balance}
\usepackage{bbm}

\usepackage{commath}
\usepackage{glossaries}
\glsdisablehyper 

\usepackage{amssymb}
\usepackage{amsfonts}
\usepackage{mathrsfs}
\usepackage{xspace}
\usepackage{bm}
\usepackage{upgreek}

\newcommand{\safemath}[2]{\newcommand{#1}{\ensuremath{#2}\xspace}}

\safemath{\bma}{\mathbf{a}}
\safemath{\bmb}{\mathbf{b}}
\safemath{\bmc}{\mathbf{c}}
\safemath{\bmd}{\mathbf{d}}
\safemath{\bme}{\mathbf{e}}
\safemath{\bmf}{\mathbf{f}}
\safemath{\bmg}{\mathbf{g}}
\safemath{\bmh}{\mathbf{h}}
\safemath{\bmi}{\mathbf{i}}
\safemath{\bmj}{\mathbf{j}}
\safemath{\bmk}{\mathbf{k}}
\safemath{\bml}{\mathbf{l}}
\safemath{\bmm}{\mathbf{m}}
\safemath{\bmn}{\mathbf{n}}
\safemath{\bmo}{\mathbf{o}}
\safemath{\bmp}{\mathbf{p}}
\safemath{\bmq}{\mathbf{q}}
\safemath{\bmr}{\mathbf{r}}
\safemath{\bms}{\mathbf{s}}
\safemath{\bmt}{\mathbf{t}}
\safemath{\bmu}{\mathbf{u}}
\safemath{\bmv}{\mathbf{v}}
\safemath{\bmw}{\mathbf{w}}
\safemath{\bmx}{\mathbf{x}}
\safemath{\bmy}{\mathbf{y}}
\safemath{\bmz}{\mathbf{z}}
\safemath{\bmzero}{\mathbf{0}}
\safemath{\bmone}{\mathbf{1}}

\safemath{\bmnu}{\mathbf{\nu}}
\safemath{\bmeta}{\mathbf{\eta}}

\bmdefine{\biad}{a}
\bmdefine{\bibd}{b}
\bmdefine{\bicd}{c}
\bmdefine{\bidd}{d}
\bmdefine{\bied}{e}
\bmdefine{\bifd}{f}
\bmdefine{\bigd}{g}
\bmdefine{\bihd}{h}
\bmdefine{\biid}{i}
\bmdefine{\bijd}{j}
\bmdefine{\bikd}{k}
\bmdefine{\bild}{l}
\bmdefine{\bimd}{m}
\bmdefine{\bind}{n}
\bmdefine{\biod}{o}
\bmdefine{\bipd}{p}
\bmdefine{\biqd}{q}
\bmdefine{\bird}{r}
\bmdefine{\bisd}{s}
\bmdefine{\bitd}{t}
\bmdefine{\biud}{u}
\bmdefine{\bivd}{v}
\bmdefine{\biwd}{w}
\bmdefine{\bixd}{x}
\bmdefine{\biyd}{y}
\bmdefine{\bizd}{z}

\bmdefine{\bixid}{\xi}
\bmdefine{\bilambdad}{\lambda}
\bmdefine{\bimud}{\mu}
\bmdefine{\bithetad}{\theta}
\bmdefine{\biphid}{\phi}
\bmdefine{\bideltad}{\delta}

\bmdefine{\birhod}{\rho}

\safemath{\bmia}{\biad}
\safemath{\bmib}{\bibd}
\safemath{\bmic}{\bicd}
\safemath{\bmid}{\bidd}
\safemath{\bmie}{\bied}
\safemath{\bmif}{\bifd}
\safemath{\bmig}{\bigd}
\safemath{\bmih}{\bihd}
\safemath{\bmii}{\biid}
\safemath{\bmij}{\bijd}
\safemath{\bmik}{\bikd}
\safemath{\bmil}{\bild}
\safemath{\bmim}{\bimd}
\safemath{\bmin}{\bind}
\safemath{\bmio}{\biod}
\safemath{\bmip}{\bipd}
\safemath{\bmiq}{\biqd}
\safemath{\bmir}{\bird}
\safemath{\bmis}{\bisd}
\safemath{\bmit}{\bitd}
\safemath{\bmiu}{\biud}
\safemath{\bmiv}{\bivd}
\safemath{\bmiw}{\biwd}
\safemath{\bmix}{\bixd}
\safemath{\bmiy}{\biyd}
\safemath{\bmiz}{\bizd}

\safemath{\bmxi}{\bixid}
\safemath{\bmlambda}{\bilambdad}
\safemath{\bmmu}{\bimud}
\safemath{\bmtheta}{\bithetad}
\safemath{\bmphi}{\biphid}
\safemath{\bmdelta}{\bideltad}

\safemath{\bmrho}{\birhod}

\safemath{\bA}{\mathbf{A}}
\safemath{\bB}{\mathbf{B}}
\safemath{\bC}{\mathbf{C}}
\safemath{\bD}{\mathbf{D}}
\safemath{\bE}{\mathbf{E}}
\safemath{\bF}{\mathbf{F}}
\safemath{\bG}{\mathbf{G}}
\safemath{\bH}{\mathbf{H}}
\safemath{\bI}{\mathbf{I}}
\safemath{\bJ}{\mathbf{J}}
\safemath{\bK}{\mathbf{K}}
\safemath{\bL}{\mathbf{L}}
\safemath{\bM}{\mathbf{M}}
\safemath{\bN}{\mathbf{N}}
\safemath{\bO}{\mathbf{O}}
\safemath{\bP}{\mathbf{P}}
\safemath{\bQ}{\mathbf{Q}}
\safemath{\bR}{\mathbf{R}}
\safemath{\bS}{\mathbf{S}}
\safemath{\bT}{\mathbf{T}}
\safemath{\bU}{\mathbf{U}}
\safemath{\bV}{\mathbf{V}}
\safemath{\bW}{\mathbf{W}}
\safemath{\bX}{\mathbf{X}}
\safemath{\bY}{\mathbf{Y}}
\safemath{\bZ}{\mathbf{Z}}

\safemath{\bZero}{\mathbf{0}}
\safemath{\bOne}{\mathbf{1}}
\safemath{\bDelta}{\mathbf{\Delta}}
\safemath{\bLambda}{\mathbf{\UpLambda}}
\safemath{\bPhi}{\mathbf{\Upphi}}
\safemath{\bSigma}{\mathbf{\Upsigma}}
\safemath{\bOmega}{\mathbf{\Upomega}}
\safemath{\bTheta}{\mathbf{\Uptheta}}

\bmdefine{\biAd}{A}
\bmdefine{\biBd}{B}
\bmdefine{\biCd}{C}
\bmdefine{\biDd}{D}
\bmdefine{\biEd}{E}
\bmdefine{\biFd}{F}
\bmdefine{\biGd}{G}
\bmdefine{\biHd}{H}
\bmdefine{\biId}{I}
\bmdefine{\biJd}{J}
\bmdefine{\biKd}{K}
\bmdefine{\biLd}{L}
\bmdefine{\biMd}{M}
\bmdefine{\biOd}{N}
\bmdefine{\biPd}{O}
\bmdefine{\biQd}{P}
\bmdefine{\biRd}{R}
\bmdefine{\biSd}{S}
\bmdefine{\biTd}{T}
\bmdefine{\biUd}{U}
\bmdefine{\biVd}{V}
\bmdefine{\biWd}{W}
\bmdefine{\biXd}{X}
\bmdefine{\biYd}{Y}
\bmdefine{\biZd}{Z}

\bmdefine{\biDelta}{\Delta}
\bmdefine{\biLambda}{\Lambda}
\bmdefine{\biPhi}{\Phi}
\bmdefine{\biSigma}{\Sigma}
\bmdefine{\biOmega}{\Omega}
\bmdefine{\biTheta}{\Theta}

\safemath{\bimA}{\biAd}
\safemath{\bimB}{\biBd}
\safemath{\bimC}{\biCd}
\safemath{\bimD}{\biDd}
\safemath{\bimE}{\biEd}
\safemath{\bimF}{\biFd}
\safemath{\bimG}{\biGd}
\safemath{\bimH}{\biHd}
\safemath{\bimI}{\biId}
\safemath{\bimJ}{\biJd}
\safemath{\bimK}{\biKd}
\safemath{\bimL}{\biLd}
\safemath{\bimM}{\biMd}
\safemath{\bimN}{\biNd}
\safemath{\bimO}{\biOd}
\safemath{\bimP}{\biPd}
\safemath{\bimQ}{\biQd}
\safemath{\bimR}{\biRd}
\safemath{\bimS}{\biSd}
\safemath{\bimT}{\biTd}
\safemath{\bimU}{\biUd}
\safemath{\bimV}{\biVd}
\safemath{\bimW}{\biWd}
\safemath{\bimX}{\biXd}
\safemath{\bimY}{\biYd}
\safemath{\bimZ}{\biZd}

\safemath{\bimDelta}{\biDelta}
\safemath{\bimLambda}{\biLambda}
\safemath{\bimPhi}{\biPhi}
\safemath{\bimSigma}{\biSigma}
\safemath{\bimOmega}{\biOmega}
\safemath{\bimTheta}{\biTheta}

\safemath{\setA}{\mathcal{A}}
\safemath{\setB}{\mathcal{B}}
\safemath{\setC}{\mathcal{C}}
\safemath{\setD}{\mathcal{D}}
\safemath{\setE}{\mathcal{E}}
\safemath{\setF}{\mathcal{F}}
\safemath{\setG}{\mathcal{G}}
\safemath{\setH}{\mathcal{H}}
\safemath{\setI}{\mathcal{I}}
\safemath{\setJ}{\mathcal{J}}
\safemath{\setK}{\mathcal{K}}
\safemath{\setL}{\mathcal{L}}
\safemath{\setM}{\mathcal{M}}
\safemath{\setN}{\mathcal{N}}
\safemath{\setO}{\mathcal{O}}
\safemath{\setP}{\mathcal{P}}
\safemath{\setQ}{\mathcal{Q}}
\safemath{\setR}{\mathcal{R}}
\safemath{\setS}{\mathcal{S}}
\safemath{\setT}{\mathcal{T}}
\safemath{\setU}{\mathcal{U}}
\safemath{\setV}{\mathcal{V}}
\safemath{\setW}{\mathcal{W}}
\safemath{\setX}{\mathcal{X}}
\safemath{\setY}{\mathcal{Y}}
\safemath{\setZ}{\mathcal{Z}}
\safemath{\emptySet}{\varnothing}

\safemath{\colA}{\mathscr{A}}
\safemath{\colB}{\mathscr{B}}
\safemath{\colC}{\mathscr{C}}
\safemath{\colD}{\mathscr{D}}
\safemath{\colE}{\mathscr{E}}
\safemath{\colF}{\mathscr{F}}
\safemath{\colG}{\mathscr{G}}
\safemath{\colH}{\mathscr{H}}
\safemath{\colI}{\mathscr{I}}
\safemath{\colJ}{\mathscr{J}}
\safemath{\colK}{\mathscr{K}}
\safemath{\colL}{\mathscr{L}}
\safemath{\colM}{\mathscr{M}}
\safemath{\colN}{\mathscr{N}}
\safemath{\colO}{\mathscr{O}}
\safemath{\colP}{\mathscr{P}}
\safemath{\colQ}{\mathscr{Q}}
\safemath{\colR}{\mathscr{R}}
\safemath{\colS}{\mathscr{S}}
\safemath{\colT}{\mathscr{T}}
\safemath{\colU}{\mathscr{U}}
\safemath{\colV}{\mathscr{V}}
\safemath{\colW}{\mathscr{W}}
\safemath{\colX}{\mathscr{X}}
\safemath{\colY}{\mathscr{Y}}
\safemath{\colZ}{\mathscr{Z}}

\safemath{\opA}{\mathbb{A}}
\safemath{\opB}{\mathbb{B}}
\safemath{\opC}{\mathbb{C}}
\safemath{\opD}{\mathbb{D}}
\safemath{\opE}{\mathbb{E}}
\safemath{\opF}{\mathbb{F}}
\safemath{\opG}{\mathbb{G}}
\safemath{\opH}{\mathbb{H}}
\safemath{\opI}{\mathbb{I}}
\safemath{\opJ}{\mathbb{J}}
\safemath{\opK}{\mathbb{K}}
\safemath{\opL}{\mathbb{L}}
\safemath{\opM}{\mathbb{M}}
\safemath{\opN}{\mathbb{N}}
\safemath{\opO}{\mathbb{O}}
\safemath{\opP}{\mathbb{P}}
\safemath{\opQ}{\mathbb{Q}}
\safemath{\opR}{\mathbb{R}}
\safemath{\opS}{\mathbb{S}}
\safemath{\opT}{\mathbb{T}}
\safemath{\opU}{\mathbb{U}}
\safemath{\opV}{\mathbb{V}}
\safemath{\opW}{\mathbb{W}}
\safemath{\opX}{\mathbb{X}}
\safemath{\opY}{\mathbb{Y}}
\safemath{\opZ}{\mathbb{Z}}
\safemath{\opZero}{\mathbb{O}}
\safemath{\identityop}{\opI}

\safemath{\veca}{\bma}
\safemath{\vecb}{\bmb}
\safemath{\vecc}{\bmc}
\safemath{\vecd}{\bmd}
\safemath{\vece}{\bme}
\safemath{\vecf}{\bmf}
\safemath{\vecg}{\bmg}
\safemath{\vech}{\bmh}
\safemath{\veci}{\bmi}
\safemath{\vecj}{\bmj}
\safemath{\veck}{\bmk}
\safemath{\vecl}{\bml}
\safemath{\vecm}{\bmm}
\safemath{\vecn}{\bmn}
\safemath{\veco}{\bmo}
\safemath{\vecp}{\bmp}
\safemath{\vecq}{\bmq}
\safemath{\vecr}{\bmr}
\safemath{\vecs}{\bms}
\safemath{\vect}{\bmt}
\safemath{\vecu}{\bmu}
\safemath{\vecv}{\bmv}
\safemath{\vecw}{\bmw}
\safemath{\vecx}{\bmx}
\safemath{\vecy}{\bmy}
\safemath{\vecz}{\bmz}

\safemath{\veczero}{\bmzero}
\safemath{\vecone}{\bmone}
\safemath{\vecxi}{\bmxi}
\safemath{\veclambda}{\bmlambda}
\safemath{\vecmu}{\bmmu}
\safemath{\vectheta}{\bmtheta}
\safemath{\vecphi}{\bmphi}
\safemath{\vecdelta}{\bmdelta}
\safemath{\vecrho}{\bmrho}
\safemath{\veceta}{\bmeta}

\safemath{\matA}{\bA}
\safemath{\matB}{\bB}
\safemath{\matC}{\bC}
\safemath{\matD}{\bD}
\safemath{\matE}{\bE}
\safemath{\matF}{\bF}
\safemath{\matG}{\bG}
\safemath{\matH}{\bH}
\safemath{\matI}{\bI}
\safemath{\matJ}{\bJ}
\safemath{\matK}{\bK}
\safemath{\matL}{\bL}
\safemath{\matM}{\bM}
\safemath{\matN}{\bN}
\safemath{\matO}{\bO}
\safemath{\matP}{\bP}
\safemath{\matQ}{\bQ}
\safemath{\matR}{\bR}
\safemath{\matS}{\bS}
\safemath{\matT}{\bT}
\safemath{\matU}{\bU}
\safemath{\matV}{\bV}
\safemath{\matW}{\bW}
\safemath{\matX}{\bX}
\safemath{\matY}{\bY}
\safemath{\matZ}{\bZ}
\safemath{\matzero}{\bmzero}

\safemath{\matDelta}{\bDelta}
\safemath{\matLambda}{\bLambda}
\safemath{\matPhi}{\bPhi}
\safemath{\matSigma}{\bSigma}
\safemath{\matOmega}{\bOmega}
\safemath{\matTheta}{\bTheta}

\safemath{\matidentity}{\matI}
\safemath{\matone}{\matO}

\safemath{\rnda}{A}
\safemath{\rndb}{B}
\safemath{\rndc}{C}
\safemath{\rndd}{D}
\safemath{\rnde}{E}
\safemath{\rndf}{F}
\safemath{\rndg}{G}
\safemath{\rndh}{H}
\safemath{\rndi}{I}
\safemath{\rndj}{J}
\safemath{\rndk}{K}
\safemath{\rndl}{L}
\safemath{\rndm}{M}
\safemath{\rndn}{N}
\safemath{\rndo}{O}
\safemath{\rndp}{P}
\safemath{\rndq}{Q}
\safemath{\rndr}{R}
\safemath{\rnds}{S}
\safemath{\rndt}{T}
\safemath{\rndu}{U}
\safemath{\rndv}{V}
\safemath{\rndw}{W}
\safemath{\rndx}{X}
\safemath{\rndy}{Y}
\safemath{\rndz}{Z}

\safemath{\rveca}{\bimA}
\safemath{\rvecb}{\bimB}
\safemath{\rvecc}{\bimC}
\safemath{\rvecd}{\bimD}
\safemath{\rvece}{\bimE}
\safemath{\rvecf}{\bimF}
\safemath{\rvecg}{\bimG}
\safemath{\rvech}{\bimH}
\safemath{\rveci}{\bimI}
\safemath{\rvecj}{\bimJ}
\safemath{\rveck}{\bimK}
\safemath{\rvecl}{\bimL}
\safemath{\rvecm}{\bimM}
\safemath{\rvecn}{\bimN}
\safemath{\rveco}{\bomO}
\safemath{\rvecp}{\bimP}
\safemath{\rvecq}{\bimQ}
\safemath{\rvecr}{\bimR}
\safemath{\rvecs}{\bimS}
\safemath{\rvect}{\bimT}
\safemath{\rvecu}{\bimU}
\safemath{\rvecv}{\bimV}
\safemath{\rvecw}{\bimW}
\safemath{\rvecx}{\bimX}
\safemath{\rvecy}{\bimY}
\safemath{\rvecz}{\bimZ}

\safemath{\rvecxi}{\bmxi}
\safemath{\rveclambda}{\bmlambda}
\safemath{\rvecmu}{\bmmu}
\safemath{\rvectheta}{\bmtheta}
\safemath{\rvecphi}{\bmphi}

\safemath{\rmatA}{\bimA}
\safemath{\rmatB}{\bimB}
\safemath{\rmatC}{\bimC}
\safemath{\rmatD}{\bimD}
\safemath{\rmatE}{\bimE}
\safemath{\rmatF}{\bimF}
\safemath{\rmatG}{\bimG}
\safemath{\rmatH}{\bimH}
\safemath{\rmatI}{\bimI}
\safemath{\rmatJ}{\bimJ}
\safemath{\rmatK}{\bimK}
\safemath{\rmatL}{\bimL}
\safemath{\rmatM}{\bimM}
\safemath{\rmatN}{\bimN}
\safemath{\rmatO}{\bimO}
\safemath{\rmatP}{\bimP}
\safemath{\rmatQ}{\bimQ}
\safemath{\rmatR}{\bimR}
\safemath{\rmatS}{\bimS}
\safemath{\rmatT}{\bimT}
\safemath{\rmatU}{\bimU}
\safemath{\rmatV}{\bimV}
\safemath{\rmatW}{\bimW}
\safemath{\rmatX}{\bimX}
\safemath{\rmatY}{\bimY}
\safemath{\rmatZ}{\bimZ}

\safemath{\rmatDelta}{\bimDelta}
\safemath{\rmatLambda}{\bimLambda}
\safemath{\rmatPhi}{\bimPhi}
\safemath{\rmatSigma}{\bimSigma}
\safemath{\rmatOmega}{\bimOmega}
\safemath{\rmatTheta}{\bimTheta}

\usepackage{amssymb}
\usepackage{amsfonts}
\usepackage{mathrsfs}
\usepackage{xspace}
\usepackage{bm}
\usepackage{fancyref}
\usepackage{textcomp}

\usepackage{multirow}
\usepackage{stmaryrd}

\newenvironment{textbmatrix}{	\setlength{\arraycolsep}{2.5pt}%
								\big[\begin{matrix}}{\end{matrix}\big]%
								\raisebox{0.08ex}{\vphantom{M}}}

\def\be{\begin{equation}}
\def\ee{\end{equation}}
\def\een{\nonumber \end{equation}}
\def\mat{\begin{bmatrix}}
\def\emat{\end{bmatrix}}
\def\btm{\begin{textbmatrix}}
\def\etm{\end{textbmatrix}}

\def\ba#1\ea{\begin{align}#1\end{align}}
\def\bas#1\eas{\begin{align*}#1\end{align*}}
\def\bs#1\es{\begin{split}#1\end{split}}
\def\bg#1\eg{\begin{gather}#1\end{gather}}
\def\bml#1\eml{\begin{multline}#1\end{multline}}
\def\bi#1\ei{\begin{itemize}#1\end{itemize}}

\DeclareMathOperator{\dg}{\opD}				
\DeclareMathOperator*{\argmin}{arg\;min}		
\DeclareMathOperator*{\argmax}{arg\;max}		
\DeclareMathOperator{\had}{\odot}			
\DeclareMathOperator{\Exop}{\opE}			

\newcommand{\conj}[1]{\ensuremath{#1^{*}}} 	
\newcommand{\tp}[1]{\ensuremath{#1^{T}}} 		
\newcommand{\herm}[1]{\ensuremath{#1^{H}}} 	
\newcommand{\inv}[1]{\ensuremath{#1^{-1}}} 	
\newcommand{\pinv}[1]{\ensuremath{#1^{\dagger}}} 	

\safemath{\dirac}{\delta}					
\safemath{\krond}{\dirac}					

\safemath{\upto}{\uparrow}
\safemath{\downto}{\downarrow}
\safemath{\iu}{j}							
\safemath{\ev}{\lambda}						
\safemath{\hilseqspace}{l^{2}}				
\newcommand{\banachfunspace}[1]{\setL^{#1}}	
\safemath{\hilfunspace}{\banachfunspace{2}}	

\safemath{\SNR}{\textit{SNR}} 				
\safemath{\PAR}{\textit{PAR}} 				
\safemath{\No}{N_0}							
\safemath{\Es}{E_s}							
\safemath{\Eb}{E_b}							
\safemath{\EbNo}{\frac{\Eb}{\No}}
\safemath{\EsNo}{\frac{\Es}{\No}}

\DeclareMathOperator{\CHop}{\ensuremath{\opH}} 
\safemath{\tvir}{\rndh_{\CHop}}				
\safemath{\tvtf}{\rndl_{\CHop}}				
\safemath{\spf}{\rnds_{\CHop}}				
\safemath{\bff}{H_{\CHop}}					

\safemath{\ircf}{r_{h}}						
\safemath{\tftvcf}{r_{s}}					
\safemath{\tfcf}{r_{l}}						
\safemath{\bfcf}{r_{H}}						

\safemath{\tcorr}{c_h}						
\safemath{\scf}{c_{s}}						
\safemath{\tfcorr}{c_{l}}					
\safemath{\fcorr}{c_{H}}						

\safemath{\mi}{I}							
\safemath{\capacity}{C}						

\safemath{\normal}{\mathcal{N}}			
\safemath{\jpg}{\mathcal{CN}}			
\safemath{\mchain}{\leftrightarrow}		

\safemath{\dB}{\,\mathrm{dB}}
\safemath{\dBm}{\,\mathrm{dBm}}
\safemath{\Hz}{\,\mathrm{Hz}}
\safemath{\kHz}{\,\mathrm{kHz}}
\safemath{\MHz}{\,\mathrm{MHz}}
\safemath{\GHz}{\,\mathrm{GHz}}
\safemath{\s}{\,\mathrm{s}}
\safemath{\ms}{\,\mathrm{ms}}
\safemath{\mus}{\,\mathrm{\text{\textmu}s}}
\safemath{\ns}{\,\mathrm{ns}}
\safemath{\ps}{\,\mathrm{ps}}
\safemath{\meter}{\,\mathrm{m}}
\safemath{\mm}{\,\mathrm{mm}}
\safemath{\cm}{\,\mathrm{cm}}
\safemath{\m}{\,\mathrm{m}}
\safemath{\W}{\,\mathrm{W}}
\safemath{\mW}{\, \mathrm{mW}}
\safemath{\J}{\,\mathrm{J}}
\safemath{\K}{\,\mathrm{K}}
\safemath{\bit}{\,\mathrm{bit}}
\safemath{\nat}{\,\mathrm{nat}}

\safemath{\define}{\triangleq}			

\safemath{\equivalent}{\sim}
\safemath{\distas}{\sim}					
\safemath{\sdiff}{\Delta}				

\safemath{\reals}{\mathbb{R}}
\safemath{\positivereals}{\reals_{+}}
\safemath{\integers}{\mathbb{Z}}
\safemath{\posint}{\integers_{+}}
\safemath{\naturals}{\mathbb{N}}
\safemath{\posnaturals}{\naturals_{+}}
\safemath{\complexset}{\mathbb{C}}
\safemath{\rationals}{\mathbb{Q}}

\newcommand*{\fancyrefapplabelprefix}{app}		
\newcommand*{\fancyrefthmlabelprefix}{thm}		
\newcommand*{\fancyreflemlabelprefix}{lem}		
\newcommand*{\fancyrefcorlabelprefix}{cor}		
\newcommand*{\fancyrefdeflabelprefix}{def}		
\newcommand*{\fancyrefproplabelprefix}{prop}		
\newcommand*{\fancyrefexmpllabelprefix}{exmpl}
\newcommand*{\fancyrefalglabelprefix}{alg}		
\newcommand*{\fancyreftbllabelprefix}{tbl}		

\frefformat{vario}{\fancyrefseclabelprefix}{Sec.~#1}
\frefformat{vario}{\fancyrefthmlabelprefix}{Theorem~#1}
\frefformat{vario}{\fancyreftbllabelprefix}{Table~#1}
\frefformat{vario}{\fancyreflemlabelprefix}{Lemma~#1}
\frefformat{vario}{\fancyrefcorlabelprefix}{Corollary~#1}
\frefformat{vario}{\fancyrefdeflabelprefix}{Definition~#1}
\frefformat{vario}{\fancyreffiglabelprefix}{Fig.~#1}
\frefformat{vario}{\fancyrefapplabelprefix}{Appendix~#1}
\frefformat{vario}{\fancyrefeqlabelprefix}{(#1)}
\frefformat{vario}{\fancyrefproplabelprefix}{Proposition~#1}
\frefformat{vario}{\fancyrefexmpllabelprefix}{Example~#1}
\frefformat{vario}{\fancyrefalglabelprefix}{Algorithm~#1}

\safemath{\dictab}{[\,\dicta\,\,\dictb\,]}

\safemath{\ysig}{\bmy}
\safemath{\ysighat}{\hat{\ysig}}
\safemath{\ysigdim}{M}
\safemath{\xsig}{\bmx}
\safemath{\xsigdim}{N}
\safemath{\nx}{n_x}
\safemath{\zsig}{\bmz}
\safemath{\zsigdim}{\ysigdim}
\safemath{\rsig}{\bmr}
\safemath{\Adict}{\bA}
\safemath{\Adicttilde}{\widetilde{\Adict}}
\safemath{\Adictdim}{\outputdim\times\xsigdim}
\safemath{\avec}{\bma}
\safemath{\avectilde}{\tilde{\avec}}
\safemath{\Bdict}{\bB}
\safemath{\Bdicttilde}{\widetilde{\Bdict}}
\safemath{\Cdict}{\bC}
\safemath{\cvec}{\bmc}
\safemath{\Ddict}{\bD}
\safemath{\Ddictdim}{\ysigdim\times\xsigdim}
\safemath{\dvec}{\bmd}
\safemath{\Ddicttilde}{\widetilde{\bD}}
\safemath{\Bonb}{\bB}
\safemath{\bvec}{\bmb}
\safemath{\Bonbdim}{\ysigdim\times\ysigdim}
\safemath{\noise}{\bmn}
\safemath{\noisedim}{\ysigim}
\safemath{\err}{\bme}
\safemath{\errdim}{\ysigdim}
\safemath{\errset}{\setE}
\safemath{\nerr}{n_e}
\safemath{\delop}{\bP_\errset}
\safemath{\delopc}{\bP_{{\errset}^c}}

\safemath{\cplxi}{\imath}
\safemath{\cplxj}{\jmath}

\safemath{\dict}{\matD}
\safemath{\inputdim}{N}		
\safemath{\outputdim}{M}		
\safemath{\sparsity}{S}	
\safemath{\inputdimA}{{N_a}}	
\safemath{\inputdimB}{{N_b}}	
\safemath{\elemA}{{n_a}}	
\safemath{\elemB}{{n_b}}	
\safemath{\resA}{\matR_a}	
\safemath{\resB}{\matR_b}	
\safemath{\subD}{\matS} 
\safemath{\subA}{\matS_a} 
\safemath{\subB}{\matS_b} 
\safemath{\dicta}{\matA} 	
\safemath{\dictb}{\matB} 	
\safemath{\hollowS}{H}
\safemath{\hollowA}{H_a}
\safemath{\hollowB}{H_b}
\safemath{\cross}{Z}
\safemath{\coh}{\mu_d}			
\safemath{\coha}{\mu_a}			
\safemath{\cohb}{\mu_b}			
\safemath{\mubs}{\nu}	
\safemath{\cohm}{\mu_m} 
\safemath{\dictset}{\setD}	
\safemath{\dictsetp}{\dictset(\coh,\coha,\cohb)}	
\safemath{\dictsetgen}{\dictset_\text{gen}}
\safemath{\dictsetgenp}{\dictsetgen(\coh)}
\safemath{\dictsetonb}{\dictset_\text{onb}}
\safemath{\dictsetonbp}{\dictsetonb(\coh)}

\safemath{\leftside}{U}
\safemath{\rightsideA}{R_a}
\safemath{\rightsideB}{R_b}

\safemath{\indexS}{\setI_S} 

\safemath{\na}{n_a}			
\safemath{\nb}{n_b}			
\safemath{\coeffa}{p_i}	
\safemath{\coeffb}{q_j}	
\safemath{\seta}{\setP}		
\safemath{\setb}{\setQ}     
\safemath{\setw}{\setW}	
\safemath{\setz}{\setZ}	
\safemath{\cola}{\veca}		
\safemath{\colb}{\vecb}		
\safemath{\cold}{\vecd}		
\safemath{\inputvec}{\vecx} 	
\safemath{\error}{\vece}	
\safemath{\noiseout}{\vecz} 	
\safemath{\inputvecel}{x}
\safemath{\inputveca}{\vecx_a}
\safemath{\inputvecb}{\vecx_b}
\safemath{\outputvec}{\vecy}	
\safemath{\lambdamin}{\lambda_{\mathrm{min}}}

\safemath{\elltwo}{\ell_2}
\safemath{\ellone}{\ell_1}
\safemath{\ellzero}{\ell_0}
\safemath{\ellinf}{\ell_\infty}
\safemath{\ellinftilde}{\ell_{\widetilde\infty}}
\safemath{\licard}{Z(\coh,\coha,\cohb)}
\safemath{\xsol}{\hat{x}}
\safemath{\xbord}{x_b}		
\safemath{\xstat}{x_s}		
\safemath{\xstatLone}{\tilde{x}_s}
\safemath{\order}{\mathcal{O}} 
\safemath{\scales}{\Theta} 
\safemath{\ones}{\mathbf{1}} 
\safemath{\zeroes}{\mathbf{0}} 
\safemath{\thlone}{\kappa(\coh,\cohb)} 
\safemath{\constoneA}{\delta} 
\safemath{\constoneB}{\epsilon} 
\safemath{\nlarge}{L}				   
\safemath{\sumlarge}{S_\nlarge}
\safemath{\maxlarger}{P_\nlarge}	   
\safemath{\Pzero}{\textrm{P0}}	
\safemath{\Pone}{\textrm{P1}}
\safemath{\vecfir}{\vecw}			 
\safemath{\vecsec}{\vecz}
\safemath{\elvecfir}{w}              
\safemath{\elvecsec}{z}				 
\safemath{\nlargefir}{n}
\safemath{\normout}{\gamma}
\safemath{\auxfun}{h}
\safemath{\supp}{\textrm{supp}}

\safemath{\indexa}{\ell}
\safemath{\indexb}{r}
\safemath{\indexc}{i}
\safemath{\indexd}{j}

\safemath{\project}{P}

\newcommand{\SoL}{\mathrm{c}}            
\newcommand{\PosVec}{\veco}
\newcommand{\PosMat}{\matO}

\newcommand{\SigData}{d}                
\newcommand{\SigCode}{c}                
\newcommand{\SigCodeVec}{\vecc}         
\newcommand{\SigTX}{t}                  
\newcommand{\SigRX}{s}                  

\newcommand{\cdt}{\SoL\,\delta t}         

\newcommand{\TD}{\Delta t}              
\newcommand{\TDInt}{\delta t}           
\newcommand{\TDist}{\tau}                  

\newcommand{\ChAtt}{\alpha}             

\newcommand{\SCAF}{\Xi}     

\newcommand{\JassProjMat}{\widetilde{\matP}}
\newcommand{\JassWindowLen}{L\sb{\SigCode}}
\newcommand{\JassPhase}{\ell}

\newcommand{\AntN}{B}                   
\newcommand{\AntI}{b}                   
\newcommand{\AntPos}{\PosVec}             
\newcommand{\AntPosAll}{\PosMat}    

\newcommand{\SatN}{S}               
\newcommand{\SatI}{\varsigma}       

\newcommand{\SPs}{\PosVec\sb{\varsigma(\nu)}}
\newcommand{\SPr}{\PosVec\sb{\varsigma(\nu')}}

\newcommand{\JassEstInterf}{\hat{I}}
\newcommand{\JamN}{I\sb{J}}             
\newcommand{\JamI}{j}                   
\newcommand{\JamChVec}{\vecj}           
\newcommand{\JamSym}{w}            
\newcommand{\JamSymVec}{\vecw}          
\newcommand{\SpfN}{I\sb{M}}             
\newcommand{\SpfI}{m}                   
\newcommand{\SpfSatN}{n_S}              
\newcommand{\SpfSet}{\zeta}             
\newcommand{\SpfChVec}{\vecq}           
\newcommand{\SpfSym}{\check{t}}         

\newcommand{\MACorrN}{Z}

\newcommand{\MSubSp}[1]{\matE\sb{#1}}

\newcommand{\MSteerVecEst}{\veca}

\newcommand{\MTh}{\tau_M}

\newcommand{\ValSetI}{\nu}              
\newcommand{\Rr}{R\sb{\varsigma(\nu')}}               
\newcommand{\Rs}{R\sb{\varsigma(\nu)}}               
\newcommand{\hatRr}{\hat{R}\sb{\varsigma(\nu')}}               
\newcommand{\hatRs}{\hat{R}\sb{\varsigma(\nu)}}               
\newcommand{\Rsr}{\Delta\sb{R}}         
\newcommand{\hatRsr}{\hat\Delta\sb{R}}         
\newcommand{\rhor}{\rho\sb{\varsigma(\nu')}}          
\newcommand{\rhos}{\rho\sb{\varsigma(\nu)}}          
\newcommand{\RotMatAnt}{\matQ}
\newcommand{\vr}{\vecv\sb{\varsigma(\nu')}}           
\newcommand{\vs}{\vecv\sb{\varsigma(\nu)}}           
\newcommand{\hatvr}{\hat\vecv\sb{\varsigma(\nu')}}           
\newcommand{\hatvs}{\hat\vecv\sb{\varsigma(\nu)}}           
\newcommand{\Prx}{\PosVec}
\newcommand{\DistEst}[2]{\hat{\rho}\sb{#1\leftarrow#2}}

\DeclareMathOperator{\diag}{\operatorname{diag}}                  
\safemath{\likef}{\mathcal{L}}

\safemath{\JSR}{\textit{JSR}}
\safemath{\SSR}{\textit{SSR}}

\newacronym{gnss}{GNSS}{global navigation satellite systems}
\newacronym{gps}{GPS}{Global Positioning System}
\newacronym{jass}{JASS}{Jammer-Aware SynchroniSation}
\newacronym{psk}{PSK}{phase-shift keying}
\newacronym{bpsk}{BPSK}{binary phase-shift keying}
\newacronym{qpsk}{QPSK}{quadrature phase-shift keying}
\newacronym{cdma}{CDMA}{code-division multiple access}
\newacronym{prn}{PRN}{pseudorandom noise}
\newacronym{sv}{SV}{space vehicle}
\newacronym{los}{LoS}{line of sight}
\newacronym{awgn}{AWGN}{additive white Gaussian noise}
\newacronym{ecef}{ECEF}{Earth-centered, Earth-fixed}
\newacronym{l1ca}{L1 C/A}{L1 Coarse Acquisition}
\newacronym{fec}{FEC}{forward error correction}
\newacronym{snr}{SNR}{signal-to-noise ratio}
\newacronym{llr}{LLR}{log-likelihood-ratio}
\newacronym{ml}{ML}{maximum likelihood}
\newacronym{mimo}{MIMO}{Multiple-Input Multiple-Output}
\newacronym{simo}{SIMO}{Single-Input Multiple-Output}
\newacronym{caf}{CAF}{cross-ambiguity function}
\newacronym{dsss}{DSSS}{direct-sequence spread spectrum}
\newacronym{cdf}{CDF}{cumulative distribution function}
\newacronym[longplural={directions of arrival}]{doa}{DoA}{direction of arrival}
\newacronym{raim}{RAIM}{receiver autonomous integrity monitoring}
\newacronym{jsr}{JSR}{jammer-to-signal ratio}
\newacronym{ssr}{SSR}{spoofer-to-signal ratio}
\newacronym{ls}{LS}{least-squares}
\newacronym{iid}{i.i.d.}{independent identically distributed}
\newacronym{irls}{IRLS}{iteratively reweighted least squares}

\newacronym{schieber}{SCHIEBER}{Spatial Cancellation of Hostile Interference by Examining Believable Emission oRigins}

\glsaddall[types={\acronymtype}]

\usepackage{framed}

\IEEEoverridecommandlockouts
\allowdisplaybreaks 

\begin{document}
\bstctlcite{IEEEexample:BSTcontrol} 

\title{GNSS Jammer and Spoofer Mitigation\\ via Multi-Antenna Processing}

\author{
	\IEEEauthorblockN{Jonas Elmiger, Gian Marti, and Christoph Studer}
	\thanks{The authors are with the Dept. of Information Technology and Electrical Engineering, ETH Zurich, Switzerland. (email: joelmiger@student.ethz.ch, marti@iis.ee.ethz.ch, studer@ethz.ch)}
	\thanks{The work of CS and GM was funded in part by an ETH Grant.}
}

\maketitle

\begin{abstract}
Modern positioning relies on radio signals from \gls{gnss}. 
Their low receive power renders these radio signals susceptible to jamming attacks, in which malicious transmitters
emit strong interference to disrupt signal acquisition. 
Moreover, \gls{gnss} are vulnerable to spoofing attacks, in which malicious transmitters mimic legitimate satellites
by transmitting spurious \gls{gnss} signals. 

We propose \glsentryshort{schieber}, a novel method for multi-antenna \gls{gnss} receivers that mitigates jammers 
as well as spoofers without requiring any prior knowledge of the receiver position or attack type: 
Jammers are mitigated during signal acquisition using a recently developed adaptive spatial 
filtering technique. Spoofers are identified and rejected after signal acquisition using a novel 
approach that tests the consistency of acquired signals by comparing their respective \gls{doa}
and pseudorange estimates in a test that is invariant with respect to the unknown receiver position. 
We demonstrate the efficacy of our method using extensive simulations of a \glsentryshort{gps} L1 C/A system under spoofing
and jamming attacks.
\end{abstract}


\section{Introduction}

\IEEEPARstart{P}{ositioning} through \gls{gnss} signals has become a vital component in countless
technologies and applications of modern life \cite{EUSPA2024MarketReport}. 
Owing to the importance of \gls{gnss} technology, numerous state actors maintain their own constellations
of navigation satellites. 
Due to the large orbital distance of these satellites, their signals arrive with significant attenuation at potential receivers. 
The low receive power of these signals and the fact that \gls{gnss} operate outdoors renders 
them vulnerable to hostile interference. In particular, \gls{gnss} are exposed to \emph{jamming attacks,} 
in which malicious transmitters try to drown \gls{gnss} signals in noise to disrupt signal acquisition at the receiver. 
Moreover, since civil \gls{gnss} use \gls{dsss} coding \cite{KaplanHegarty2017GNSS} with publicly known spreading codes,
they are also vulnerable to \emph{spoofing attacks,} in which malicious transmitters mimic legitimate satellites by emitting
spurious signals that are encoded with valid \gls{gnss} spreading codes. 
Such spoofing can take various forms \cite{Wu2020SurveySpoofASpf}: While simple spoofing attacks cause the receiver to fail
in correctly positioning itself, more sophisticated attacks such as replaying or meaconing
\cite{papadimitratos2008gnssAttacks} can take over the estimated position and steer it away from the true location. 
The use of encrypted spreading codes can alleviate the risk of spoofing but necessitates key confidentiality and is typically only 
used in military signals. 
A possible defense mechanism against the spoofing of civil signals is message integrity protection as implemented 
by Galileo Open Service Navigation Message Authentication (OS-NMA) \cite{Fern2016GalileoOSNMA, EUSPA2024OSNMA}. 
OS-NMA authenticates transmission data based on early message signing with delayed key disclosure. 
In itself, however, it only permits detection---not mitigation---of spoofing attacks.

Multi-antenna receivers have the ability to spatially resolve signals, 
which allows them to (i) distinguish between signals from different origins and (ii) use spatial filtering
for nulling signals from certain directions---this makes multi-antenna processing a promising basis for jammer and spoofer mitigation.

\subsection{Contributions} 
We propose \glsentryshort{schieber}, \glsunset{schieber} 
a novel method for jammer and spoofer mitigation in multi-antenna \gls{gnss} receivers. 
\gls{schieber} can mitigate simultaneous attacks by multiple jammers and spoofers 
without requiring encrypted spreading codes and without requiring any \emph{a priori} estimate 
of the receiver position. All that is required is a local almanac of current satellite positions as well as a rough estimate of the time, 
so that the receiver can autonomously infer the current positions of the \gls{gnss} satellites.
\gls{schieber} mitigates jammers during signal acquisition while spoofers are mitigated between pseudorange estimation 
and positioning.  

For signal acquisition, \gls{schieber} builds on a recently developed method for jammer-resilient synchronization 
\cite{marti2025sync}, which uses adaptive spatial filtering for mitigating signals that interfere with the synchronization
(or spreading) code. This enables \gls{schieber} to acquire satellite signals even under strong multi-antenna jamming while 
simultaneously maintaining the basic principles of regular GNSS signal acquisition. 
Since the signal acquisition stage does not distinguish between spoofers and legitimate satellites, spoofed
signals are identified and rejected in a second stage. To this end, \gls{schieber} uses a novel approach 
in which it tests the consistency of \emph{pairs} of acquired signals by comparing their respective
directions of arrival (DoAs) and pseudorange estimates in a test that is invariant with respect to the 
unknown receiver position. The DoAs themselves are estimated using a modified version of the multiple signal classification (MUSIC) \cite{schmidt1986music} algorithm that takes into account the spatial filter from the jammer-mitigating 
signal acquisition stage. 
Finally, to increase the robustness in case that a spoofed signal has slipped through the spoofer rejection stage, 
\gls{schieber} uses an outlier-robust \gls{irls} approach to estimate the position based on those pseudoranges
that were not rejected. 
We demonstrate the efficacy of \gls{schieber} through extensive simulations of a \gls{gps} L1 C/A system \cite{GPS_L1}
under various spoofing and jamming attacks.

Our approach aspires to refrain from making unreasonable assumptions about the attackers. 
In particular, we do not assume that the number of jammers or spoofers is known, 
and we do not make assumptions on the strength of their signals, 
nor about how many satellites a single spoofer spoofs simultaneously.

\subsection{Related Work}
Through its use of \gls{dsss}, \gls{gnss} is inherently designed to be robust to interference
\cite{hegarty2012gnssSignals}. Moreover, time-frequency filtering can further improve the resilience
against interference with little additional effort \cite{milstein1988dsssInterf}. However, 
such mitigation techniques are effective mainly against narrowband jammers. 
The same holds true for the adaptive notch filtering techniques proposed by \cite{borio2021SpoofingMitig}. 
However, while these methods are easy to integrate into traditional single-antenna receivers, 
they are ineffective against strong wideband jamming attacks \cite{wu2018interfMitigGNSS}. 
One possibility for mitigating such attacks is given by multi-antenna receivers, which can block hostile
interference through spatial filtering (also known as beamforming): Reference~\cite{Amin2005SCORE}
leverages the beamforming capabilities of multi-antenna receivers by exploiting the spreading code repetitions 
in \gls{gps} L1 C/A, which are used to form beams towards legitimate satellites. 
However, the assumption of perfect Doppler tracking limits this method to a single legitimate signal.
Alternatively, the method from~\cite{wu2018interfMitigGNSS} minimizes the power after the spatial filter under 
the assumption that the strongest signal directions correspond to jamming and/or noise. 
Similarly, the method from~\cite{daneshmand2013interfSuppr} treats the leading eigenvectors of the receive signal's 
spatial covariance matrix as estimates of the jamming signal, which are then null-steered. 
A main drawback of such techniques is that they require assumptions about how many interference dimensions 
need to be nulled (i.e., how many jammers are jamming), and that they can fail when that assumption is violated. 
Moreover, such methods fail against interference that mimics legitimate satellite signals, i.e., against spoofing.

Spoofer mitigation is often performed via completely different methods than jammer mitigation. 
The commonly implemented \gls{raim} and Advanced \gls{raim} \cite{betz2016nav} methods exploit the redundancy 
when receiving five or more satellites to find outliers among the signals. 
While these methods can detect spoofing events caused by single spoofers, they are (i) designed mainly to ensure satellite integrity and (ii) are vulnerable to spoofing through multiple spoofers.
A method for general spoofing detection with single-antenna receivers is proposed in~\cite{Rang2016SPREE}, 
where a takeover attack that aims to influence the derived position is assumed. Using a correct initial position 
in combination with tracking, their approach raises a spoofing alarm upon detecting multiple peaks during the acquisition 
of a single satellite spreading code.  In similar fashion, reference~\cite{Han2017SpoofMitig} proposes to identify spoofing by noticing when a single satellite
spreading code leads to multiple acquired signals, and to reject the signals that were acquired earlier (based on an 
assumption that spoofed signals will be acquired first because the spoofer signal is assumed to be stronger than
the legitimate signal). 
Another alternative is described in \cite{Gulgun2021SpoofDetectStatistics}, where the assumption of a 
straight-line receiver trajectory is used. This approach enables one to examine changes in signal Doppler and phase, 
which must be different between signals that originate from different signal sources. 
If a single spoofer spoofs multiple signals, it can be detected this way. 
Just as for jammer mitigation, the spatial resolution capabilities of multi-antenna receivers are also 
useful for spoofer mitigation. Methods for multi-antenna spoofing detection are proposed 
in~\cite{Psiaki2014SpoofDetect2antenna} and \cite{Rothmaier2021SpoofDetectSpatProc}, which exploit that dual-antenna 
spatial diversity must create varying carrier-phase differences among legitimate signals. Since these characteristics are 
dependent on the receive \gls{doa}, an attacker cannot replicate the diversity characteristics when spoofing multiple
satellite signals simultaneously from the same direction. 
The same assumption (of a single attacker antenna spoofing multiple satellite signals) is used in~\cite{Zhao2022CoprimeArrayDoA}, which detects spoofing and estimates its \gls{doa} using a coprime-array. 
Direct \gls{doa} examination is performed in \cite{Meurer2012SpoofDetectDoAAttiEstim}, where subsets of the measured \glspl{doa} are iteratively used to estimate the receiver's attitude and evaluated for compatibility 
with the expected \glspl{doa}. While this approach is not dependent on an assumption that a single spoofer spoofs 
multiple satellites, its computational complexity increases exponentially with the number of acquired signals. 
Reference \cite{Shi2019MultiParticipantDoAVerific} similarly suggests a \gls{doa} examination approach that compares 
\gls{doa} measurements against the almanac information. The considered distributed sensor array consists of 
an aircraft fleet with single-antenna receivers, which estimates the directions from signal arrival time differences. 
The required assumptions of perfectly synchronized clocks among receivers and precise awareness of their relative positions to each other are optimistic, which speaks in favor a multi-antenna single-platform approach.

Combined jamming and spoofing mitigation is demonstrated in \cite{Hu2018JamSpoofMitig, Zhang2019JamSpoofSuppress},
which extract the jamming subspace from the incoming signal's spatial covariance matrix by assuming that the jamming 
and noise components dominate, and then project the signal into the orthogonal complement space. 
Their accompanying spoofing suppression is again based on an assumption that a single attacker imitates multiple signals simultaneously from the same direction. This allows the elimination of the strongest spatial direction once it 
crosses a spoofing detection threshold. 
Reference \cite{yang2019JamSpoofAntiAlgo} presents the same jamming suppression approach and likewise assumes a 
single-antenna multi-satellite spoofer, but explicitly estimates \gls{doa} angles to eliminate signal directions appearing more than once.
A similar combined approach is explained in \cite{zorn2018JamSpoofPosAtt}, where the same jamming elimination projection 
is applied, but where another orthogonal projection in each signal tracking stage is added. This  latter projection matrix is 
estimated via an antenna array orientation estimation, which is tracked by a Kalman filter and which allows conflicting signal 
directions to be found. This approach requires a rough initial position estimate to determine the true satellite \glspl{doa}.

In summary, while separate jammer and spoofer mitigation strategies can often be combined, 
only few works explicitly demonstrate the simultaneous efficacy of both. Moreover, many mitigation 
approaches are dependent on specific assumptions about the attackers (such as that the number of jammers is known, 
that a single spoofer spoofs multiple satellites simultaneously, or that the jammers or spoofer signals are 
much stronger than legitimate satellite signals), or they depend on an assumption that the receiver
approximately knows its position already from the beginning. 
In contrast to our method, none of these previous works present a feasible approach for positioning under both 
multi-antenna jamming and multi-antenna spoofing \emph{without} any prior position information. 

\subsection{Notation} \label{sec:notation}
We use boldface lowercase and uppercase letters to denote column vectors 
\mbox{$\bma=\tp{[a_1,\dots,a_n]}\!=[a_1;\dots;a_n]\!\in\opC^n$}
and matrices $\bA=[\bma_1,\dots,\bma_m]\in\opC^{n\times m}$, respectively. 
Complex conjugation, transposition, conjugate transposition, and Moore-Penrose pseudo-inversion are denoted as
$\conj{\bA}$, $\tp{\bA}$, $\herm{\bA}$, and~$\pinv{\bA}$, respectively. 
The Hadamard product of $\bA$ and $\bB$ is denoted $\bA\had\bB$. 
The Euclidean norm of $\bma$ is $\|\bma\|_2$, and the Frobenius norm of $\bA$ is $\|\bA\|_F$. 
The $n\times n$ identity matrix is $\bI_n$, and the all-zero vector is $\boldsymbol{0}_n\in\opC^n$ 
(depending on the context, the subscript $n$ indicating the dimension may be omitted). 
The cardinality of a set $\setA$ is $|\setA|$, the largest integer smaller than or equal to $x$ is $\lfloor x\rfloor$, 
and the distribution of a circularly symmetric complex Gaussian random vector with covariance matrix $\bC$ 
is $\setC\setN(\boldsymbol{0},\bC)$. 
Estimates are usually marked by a hat (e.g., $\hat{x}$ is an estimate of $x$), and values that are iteratively 
updated are marked by a bracketed superscript (e.g., $x^{(k)}$, where $k$ is the iteration index). 

\section{Model and Prerequisites} \label{sec:prerequisites}

\subsection{Signal Model} \label{sec:signal_model}
We consider a stationary receiver equipped with $B$ antennas located at some (unknown) position $\PosVec=\tp{[x,y,z]}$
(expressed in  \gls{ecef} coordinates), and a satellite constellation consisting of $S$ active satellites 
with instantaneous positions $\PosVec_\SatI,\SatI=1,\dots,S$. 
We assume a so-called \emph{warm start}~\cite{betz2016nav}, which means that the receiver knows the instantaneous time
(up to an unknown internal clock offset~$\TDInt$) and possesses an almanac of the satellite trajectories, 
so that it can infer the instantaneous positions $\PosVec_\SatI$ of all satellites $\SatI=1,\dots,S$.

The baseband receive signal at sample index $k$ under the impact of $\JamN\geq0$ stationary single-antenna jammers 
and \mbox{$\SpfN\geq0$} stationary single-antenna spoofers is modeled as
\begin{align} \label{eq:io_model}
    \vecy[k] =& \sum\sb{\SatI=1}\sp{\SatN}\vech\sb{\SatI}\![k]\, \SigTX\sb{\SatI}\sbr{k-\Big\lfloor\frac{\TD\sb{\SatI}}{T}\Big\rfloor} \nonumber \\
    &+ \sum\sb{\JamI=1}\sp{\JamN}\JamChVec\sb{\JamI} \JamSym\sb{\JamI}\sbr{k-\Big\lfloor\frac{\TD\sb{\JamI}}{T}\Big\rfloor} \nonumber \\ 
    &+ \sum\sb{\SpfI=1}\sp{\SpfN}\SpfChVec\sb{\SpfI} \SpfSym\sb{\SpfI}\sbr{k-\Big\lfloor\frac{\TD\sb{\SpfI}}{T}\Big\rfloor} + \vecn[k],
\end{align}
where $\bmh_\SatI[\cdot]\in\opC^B$ is the channel from the $\SatI$th satellite to the receiver, 
$\SigTX_\SatI[\cdot]\in\opC$ is the $\SatI$th satellite's transmit signal, $\TD\sb{\SatI}$ is the signal delay between 
the $\SatI$th satellite and the receiver, and $T$ is the sampling period. 
Similarly, $\JamChVec\sb{\JamI}\in\opC^{B}$ and $\SpfChVec\sb{\SpfI}\in\opC^{B}$ are the channels of the $\JamI$th jammer and 
the $\SpfI$th spoofer, respectively; $\JamSym\sb{\JamI}[\cdot]$ and $\SpfSym\sb{\SpfI}[\cdot]$ are their respective transmit signals;
and $\TD_\JamI$ and $\TD_\SpfI$ are their respective path delays. 
Finally, $\vecn[k] \sim \mathcal{CN}(\veczero,\sigma\sb{n}\sp{2}\matI\sb{\AntN})$ is \gls{awgn}. 

\begin{figure}
    \centering
    \includegraphics[width=0.8\linewidth]{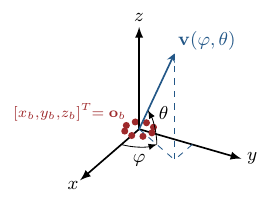}
    \vspace{-2mm}
    \caption{Illustration of a unit vector $\bmv(\varphi,\theta)$ in a receiver-centric coordinate system 
    with $B=8$ receive antennas that are arranged in a ring in the \mbox{$xy$-plane}
    (with the $b$th antenna located at cartesian coordinates
    $\AntPos\sb{\AntI} = \tp{\sbr{x\sb{\AntI},\,y\sb{\AntI},\,z\sb{\AntI}}}$).}
    \label{fig:ray_onto_antennas}
\end{figure}

We assume that the channels between the satellites and the receiver are \gls{los} with negligible multipath. 
Hence, the channel from the $\SatI$th satellite to the receiver at sample index $k$ can be expressed as
\begin{align} \label{eq:los_channel}
    \vech\sb{\SatI}[k] = \ChAtt\sb{\SatI} e\sp{i2\pi f\sb{\SatI} k T}\veca\sb{\SatI},
\end{align}
where $\ChAtt\sb{\SatI}\in\complexset$ is the distance-induced complex attenuation, which is equal to zero if the $\SatI$th 
satellite is behind the horizon, $f_\SatI$ is the Doppler shift incurred by satellite motion (assumed to be constant over the 
considered timescales), 
and $\veca\sb{\SatI}\in\complexset\sp{\AntN}$ is the antenna arrangement steering vector 
(also assumed constant over the considered timescales), which is given as
\begin{align} \label{eq:steering_vector}
    \veca\sb{\SatI} = \exp\del{-i2\pi \tp{\AntPosAll}\vecv(\theta\sb{\SatI},\varphi\sb{\SatI}) /\lambda}\!,
\end{align}
where the columns $\AntPos\sb{\AntI} = \tp{\sbr{x\sb{\AntI},\,y\sb{\AntI},\,z\sb{\AntI}}}$
of $\AntPosAll = \sbr{\AntPos\sb{1},\,\ldots,\,\AntPos\sb{\AntN}}$
contain the position coordinates of the receive antennas in a receiver-centric Cartesian coordinate system, and 
where $\vecv(\theta,\varphi) = \tp{\sbr{\cos{\theta}\cos{\varphi},\,\cos{\theta}\sin{\varphi},\,\sin{\theta}}}$
is a unit vector pointing in the direction of the receive signal with elevation~$\theta$ and azimuth~$\varphi$ 
(cf. \fref{fig:ray_onto_antennas}), and $\lambda$ is the signal wavelength. 

The signal delay $\TD\sb{\SatI}=\TDist\sb{\SatI}+\TDInt$ is comprised of the path delay 
$\TDist\sb{\SatI}=\|\PosVec_\SatI-\PosVec\|_2/c$ (where $c$ is the speed of light) 
and the receiver's internal clock offset $\TDInt$ from the global reference clock. 
This quantity represents the message delay perceived by the receiver with respect to its local clock sampling origin $k=0$. 
For simplicity, we omit additional delay factors such as satellite-internal offsets or atmospheric influences.

As stated above, the receiver, the jammers, and the spoofers are assumed to be stationary, which is why the channels~$\JamChVec_\JamI$ and $\SpfChVec_\SpfI$ do not depend on $k$. 
However, our methods are straightforwardly applicable also in situations
where---like the satellites---these transceivers are moving with constant velocity. Moreover, we assume that the spoofers
emulate satellite-like motion by modulating their transmit signals with virtual Doppler shifts (cf. \fref{sec:interference_model}). 
We assume that the jammers and the spoofers either also exhibit \gls{los} characteristics, in which case 
they have the form $\JamChVec_\JamI=\ChAtt_\JamI\bma_\JamI$ and $\SpfChVec_\SpfI=\ChAtt_\SpfI\bma_\SpfI$
with steering vectors as defined in~\eqref{eq:steering_vector}, or that they exhibit \gls{iid} Rayleigh fading characteristics.

The $\SatI$th satellite's unit-power transmit signal is
\begin{align} \label{eq:tx_symbol}
    \SigTX\sb{\SatI}[k] = \SigData\sb{\SatI}\big[\lfloor k/\JassWindowLen\rfloor\big] \SigCode\sb{\SatI}[k\text{ mod }\JassWindowLen ]
\end{align}
and consists of the product of the satellite's spreading code $\bmc_\SatI=\big[c_\SatI[0];\dots;c_\SatI[\JassWindowLen-1]\big]$ 
and its data symbols $\SigData\sb{\SatI}[\cdot]$. In real-world GNSS systems, these data symbols 
contain information about the current time, the satellite trajectory, satellite clock corrections, 
and more. In this work, however, we assume that the receiver knows the current time up to the timing error~$\TDInt$ and has an offline almanac available, which allows it to estimate the current position $\PosVec\sb{\SatI}$ 
of each satellite (this is called a warm start \cite{betz2016nav}) without relying on the data symbols 
$\SigData\sb{\SatI}[\cdot]$. Hence, the only function of the data symbols is to 
impart the receiver with information about the path delay. To this end, we assume the data symbols to be given as
\begin{align}
	\SigData\sb{\SatI}[K] = \begin{cases} -1 & K<K_0 \\ +1 & K\geq K_0
 \end{cases}
 \label{eq:data_sequence}
\end{align}
for some publicly known instant $K_0$, which can be seen as a simplified model of a message header.

\subsection{Attack Model} \label{sec:interference_model}
We distinguish between to types of interference: 

\emph{Jamming} describes interference consisting of noise-like random symbols. 
Throughout this work, when considering single-jammer attack scenarios, we model the jammer transmit symbols as
zero-mean, \gls{iid} complex random variables (also known as barrage jamming)
\begin{align} \label{eq:jammer_single_ant}
    \JamSym[k] \sim \mathcal{CN}(0,N\sb{J}).
\end{align}
When considering distributed or multi-antenna jammer scenarios, we model the jammers as dynamically beamforming jammers, 
which are more challenging to mitigate than jammers that use constant beamforming \cite{marti2023universal}.
For jammers with a total of~$\JamN$ transmit antennas, the jammer transmit vectors are modeled as
\begin{align} 
    \JamSymVec[k] & = \matB[k] \vecb[k] \\
    \matB[k+1] & = \begin{cases}
        \matB[k] & \text{w.p. } 95\% \\
        \Pi \diag\del{\tp{\vecone_R},\tp{\veczero\sb{\JamN-R}}} & \text{w.p. } 5\%,
    \end{cases} \label{eq:jammer_multi_ant}
\end{align}
where $\vecb[\cdot]\stackrel{\text{i.i.d.}}{\sim}\mathcal{CN}(\veczero,\sigma\sb{J}^2\matI\sb{\JamN})$ are jamming symbol vectors and $\Pi$ is a random permutation matrix. Essentially the jamming symbols are beamformed using a matrix $\matB\in\complexset\sp{\AntN\times\JamN}$ which activates $R$ jamming antennas simultaneously, deciding randomly which $R$ of the $\JamN$ antennas to use.
We note, however, that the specific transmission model of the jammers is not critical for the efficacy of our 
mitigation method (e.g., non-Gaussian transmit signals would also be mitigated). The only relevant assumption is 
that the jammer transmit signals are random and independent of the satellite signals. 

\emph{Spoofing,} in contrast, is used to describe the intentional imitation of spreading codes or full satellite signals. 
While aimless, intermittent imitation of spreading codes without the transmission of sensible data symbols as in 
\eqref{eq:tx_symbol} can disrupt the regular acquisition approach, the subsequent decoding either produces insensible data 
or fails outright. In contrast, the imitation of entire satellite data streams, either through replaying received signals 
(meaconing) or through emulation of satellite behavior, is harder to recognize and reject from positioning. 
We therefore consider smart spoofers that actually transmit satellite-like data symbols as in \eqref{eq:data_sequence}. 
Hence, the transmit signal of a spoofer $m\in\{1,\dots,\SpfN\}$ is\footnote{In contrast to jamming, 
our spoofer transmit model does distinguish between single spoofer scenario and multi-spoofer scenarios.}
\begin{align} \label{eq:spoofer_tx_symbol}
    \SpfSym\sb{\SpfI}[k] = \check{a}_m\sum\sb{\varsigma'\in\SpfSet_m}\check{\SigData}\sb{\varsigma'}[k+\delta t\sb{\varsigma'}]\SigCode\sb{\varsigma'}[k+\delta t\sb{\varsigma'}] e\sp{i2\pi f\sb{\varsigma'}kT}.
\end{align}
Here, the spoofer imitates satellites from an index set $\SpfSet_m$ which contains the indices of $\SpfSatN=|\SpfSet_m|$ 
satellites to be spoofed. The data symbols $\check{\SigData}_{\SatI'}[\cdot]\in\complexset$ produce a credible message 
and are spread with the corresponding spreading codes of the mimicked satellites. A spoofer emulates the 
Doppler effect $f\sb{\varsigma'}$ as attackers are stationary in our model, and it introduces an individual delay 
$\delta t\sb{\varsigma'}$ for each spoofed satellite. 
This allows an adversary to create believable constellation streams in the receiver (apart from the different \glspl{doa},
which cannot be emulated by a single spoofer).
Finally, $\check{a}_m$ is a parameter for controlling the spoofer's signal power.
Spoofing attacks attempting to cancel the original signal (i.e., to null it at the receiver) \cite{psiaki2016gnssSpoofingAttacks} require targeting and are extremely challenging to execute. Such an approach would require precise knowledge of the receiver's position, the attacker's distance to the receiver, and must match the original channel's features such that it is hardly feasible to implement against multi-antenna receivers. Thus, we have chosen not to consider it in our attacker model.

\subsection{GNSS Prerequisites} \label{sec:gnss_basics}
We now provide a quick outline of regular GNSS positioning (i.e., GNSS positioning in the absence of jamming and spoofing) 
for a multi-antenna receiver
that will serve as background for our method in \fref{sec:methods}, and as the baseline for our evaluation in \fref{sec:eval}. 
We distinguish three stages: signal acquisition, pseudorange estimation, and positioning. 

\subsubsection{Signal Acquisition} \label{sec:acquisition}
Even in the absence of jamming and spoofing (i.e., when $\JamN=\SpfN=0$), the receive signal in~\eqref{eq:io_model} contains
the time-delayed superposition of many satellite signals, all of which are considerably below the noise floor. 
When acquiring the $\SatI$th satellite's signal, the receiver searches for the code phase 
$\JassPhase_\SatI\triangleq \Big\lfloor\frac{\TD\sb{\SatI}}{T}\Big\rfloor\text{ mod }\JassWindowLen$
(as viewed from the receiver) of the spreading code 
as well as the corresponding Doppler shift~$f\sb{\SatI}$; see~\eqref{eq:los_channel}. 
To this end, the receiver uses the 
$\SatI$th \gls{caf} \cite{betz2016nav} which, for our multi-antenna receiver, we define as
\begin{align} 
    \SCAF\sb{\SatI}[\JassPhase,f] &= \frac{\norm{\sum\sb{k=0}\sp{\JassWindowLen-1} \vecy[k+\JassPhase] \SigCode\sb{\SatI}\sp{\ast}[k] e\sp{-i2\pi f kT} }_2^2}{\sum\sb{k=0}\sp{\JassWindowLen-1}\norm{\vecy[k+\JassPhase]}_2^2} \label{eq:regular_caf}\\
    &= \frac{\norm{\matY[\JassPhase] \dg(f) \SigCodeVec\sb{\SatI}\sp{\ast} }_2^2}{\norm{\matY[\JassPhase]}_F^2}, \label{eq:matrix_caf}
\end{align}
where 
\begin{align}
    \matY[\JassPhase] &\triangleq \sbr{\vecy[\JassPhase],\ldots,\,\vecy[\JassPhase+\JassWindowLen-1]} \label{eq:y_window}
\end{align}
is the windowed receive signal and
\begin{align}
    \dg(f) &= \diag(1,\,e\sp{-i2\pi f T},\,\ldots,\,e\sp{-i2\pi f\del{\JassWindowLen-1}T}) \label{eq:dopp_diag}
\end{align}
is a diagonal matrix to compensate the Doppler shift. 
The $\SatI$th CAF measures the Doppler-adjusted and energy-normalized correlation of the windowed receive signal with the $\SatI$th 
satellite's sprading code $\SigCodeVec_\SatI$.\footnote{In principle, the length over which the CAF measures the correlation 
may differ from the length of the spreading code. For simplicity, however, we assume these two lengths to be equal.}
For the acquisition of the $\SatI$th satellite's signal, the receiver searches for the maximum of the \gls{caf}
over all possible code phases $\JassPhase\in\{1,\ldots,\JassWindowLen\}$ as well as over a quantized grid of potential Doppler shift compensations~$f\in\setF$: 
\begin{align} \label{eq:normal_acquisition}
    (\hat\JassPhase\sb{\SatI},\hat{f}\sb{\SatI})=\underset{\substack{\JassPhase\in\{1,\ldots,\JassWindowLen\}\\ f\in\setF}}{\argmax}\cbr{
    \SCAF\sb{\SatI}[\JassPhase,f]
    \;\middle|\;
    \SCAF\sb{\SatI}[\JassPhase,f] \geq\tau }\!.
\end{align}
Note that, to prevent bogus acquisitions of Earth-obstructed satellites, 
the receiver only acquires a signal if the maximum of the \gls{caf} exceeds some threshold $\tau$.

\subsubsection{Pseudorange Estimation} \label{sec:decode}
When a signal has been acquired at phase $\hat\JassPhase_\SatI$ and Doppler shift $\hat{f}_\SatI$, it is despreaded
using a replica of the $\SatI$th satellite's spreading code~$\SigCodeVec\sb{\SatI}$, resulting in the symbol vector $\bmr_\SatI$. 
Defining the windowed transmit signal $\tp{\bmt}_\SatI[\ell]=\big[t_\SatI[\ell], \dots, t_\SatI[\ell+\JassWindowLen-1]\big]$, and
assuming a perfect match of the acquired Doppler shift $\hat{f}_\SatI=f_\SatI$ 
and code phase $\hat\JassPhase_\SatI\equiv \Big\lfloor\frac{\TD\sb{\SatI}}{T}\Big\rfloor\text{ mod }\JassWindowLen$,
the $K$th symbol vector is computed as
\begin{align} \label{eq:symbol_vector_despread}
    &\vecr\sb{\SatI}[K] \nonumber \\
    & \triangleq 
     \matY[K\JassWindowLen + \hat\JassPhase\sb{\SatI}]\dg(\hat{f}_\SatI) \SigCodeVec\sb{\SatI}\sp{\ast} \\
    &= \alpha\sb{\SatI}\veca\sb{\SatI} \tp{\bmt_\SatI}\bigg[K\JassWindowLen + \underbrace{\hat\JassPhase\sb{\SatI} - \Big\lfloor\frac{\TD\sb{\SatI}}{T}\Big\rfloor}_{\triangleq \Delta K_\SatI \JassWindowLen}\bigg] \SigCodeVec\sb{\SatI}\sp{\ast} \nonumber \\
    &\hphantom{=}+\! \sum\sb{\SatI'\neq\SatI}\alpha\sb{\SatI'}\bma_{\SatI'} \dg(\hat{f}\sb{\SatI}\!-\!f_{\SatI'}) \tp{\bmt_{\SatI'}}\bigg[K\JassWindowLen + \hat\JassPhase\sb{\SatI'} \!-\! \Big\lfloor\frac{\TD\sb{\SatI'}}{T}\Big\rfloor\bigg]  \SigCodeVec\sb{\SatI}\sp{\ast} \nonumber \\
    &\hphantom{=}\underbrace{+\matN\dg(\hat{f}\sb{\SatI}) \SigCodeVec\sb{\SatI}\sp{\ast}\hspace{6cm}}_{\triangleq \tilde\bmn}
    \label{eq:other_satellite_interference}\\
    &= \alpha\sb{\SatI}\,\veca\sb{\SatI} \tp{\bmt_\SatI}\left[(K - \Delta K_\SatI)\JassWindowLen\right] \SigCodeVec\sb{\SatI}\sp{\ast} + \tilde\bmn \label{eq:mod_Lc_equality} \\
    &= \alpha\sb{\SatI}\,\veca\sb{\SatI}d\sb{\SatI}[K-\Delta K\sb{\SatI}] \norm{\SigCodeVec\sb{\SatI}}^2 + \tilde\bmn \label{eq:use_tx_symbol}
\end{align}
where $\Delta K$ is the data index retardation by which  the satellite's despreaded data symbol 
$D\sb{\SatI}[\cdot]$ appears delayed due to the path delay, where $\tilde\bmn\in\opC^B$ subsumes 
residual disturbance from the other satellite signals and thermal noise, 
where \eqref{eq:mod_Lc_equality} follows from the fact 
that $\hat\JassPhase_\SatI\equiv \Big\lfloor\frac{\TD\sb{\SatI}}{T}\Big\rfloor\text{ mod }\JassWindowLen$, 
and where~\eqref{eq:use_tx_symbol} follows from \eqref{eq:tx_symbol}.
Since the other spreading codes $\SigCodeVec_{\SatI'}$ are only weakly correlated with $\SigCodeVec_\SatI$ regardless 
of whether or not they are cyclically shifted, the residual disturbance terms from the other satellites 
in \eqref{eq:other_satellite_interference} are significantly attenuated compared to the despreaded signal 
from the $\SatI$th satellite.

Note that $\bma_\SatI$ is not known at the receiver because the receiver knows neither its current position 
nor its orientation. However, the receiver can estimate the data retardation $\Delta K$ from the symbol vector sequence
$\bmr_\SatI[\cdot]$ by searching for the $\hat{K}_\SatI$, for which $\bmr_\SatI[\hat{K}_\SatI]$ exhibits the phase change
corresponding to $K_0$ in \eqref{eq:data_sequence}, and estimating $\Delta \hat{K}_\SatI=\hat{K}_\SatI-K_0$.
Using the estimated data retardation $\Delta \hat{K}_\SatI$ and the code phase $\hat\JassPhase\sb{\SatI}$, 
the receiver can then estimate the sum of the path delay and its internal time offset 
$\TD_\JamI = \TDist\sb{\SatI} + \TDInt$ as $(\hat\JassPhase\sb{\SatI}+\JassWindowLen\,\Delta \hat{K}\sb{\SatI})T$. 
Multiplication with the speed of light $c$ gives the so-called \emph{pseudorange} estimate
\begin{align} \label{eq:est_pseudorange}
	\hat{R}_\SatI \triangleq c(\hat\JassPhase\sb{\SatI}+\JassWindowLen\,\Delta \hat{K}\sb{\SatI})T.
\end{align}

\subsubsection{Positioning} \label{sec:positioning}
Barring non-idealities, the ``true'' pseudorange $R_\SatI$ is a function of the coordinates 
$\PosVec_\SatI=\tp{[x_\SatI, y_\SatI, z_\SatI]}$ and $\PosVec=\tp{[x,y,z]}$ 
of the $\SatI$th satellite and the receiver, respectively. 
The pseudorange is then given as 
\begin{align} \label{eq:pseudorange_exact}
	R_\SatI = \underbrace{\|\PosVec_\SatI-\PosVec\|_2}_{\triangleq \rho(\PosVec_\SatI,\PosVec)\hspace{-0.9cm}} + \cdt. 
\end{align}
That is, $R_\SatI$ is a function of the four unknown variables $x,y,z,$ and $\cdt$ 
and the four known variables $x_\SatI,y_\SatI,$ and $z_\SatI$. 
Denote the set of satellites $\SatI$ whose pseudoranges have been sucessfully acquired by $\setS$, 
then the receiver position can be inferred (provided that $|\setS|\geq4$) by solving the least squares problem
\begin{align}
	\min_{\PosVec,\cdt} \, \sum_{\SatI \in \setS} \left( \hat{R}_\SatI  - \rho(\PosVec_\SatI,\PosVec) - \cdt \right)^2\!, 
	\label{eq:gnss_pos_opt}
\end{align}
where $\hat{R}_\SatI$ is the estimate of $R_\SatI$ from \eqref{eq:est_pseudorange}.
A difficulty lies in the nonlinearity of $\rho(\PosVec_\SatI,\PosVec)$, which is commonly addressed by using 
an iterative approach in which $\rho(\PosVec_\SatI,\PosVec)$ is linearized at every iteration 
\cite[Sec.~2.5.2]{KaplanHegarty2017GNSS}: if 
$\PosVec^{(k)}$ is the estimated receiver position at iteration $k$, then the first-order Taylor approximation of 
$\rho(\PosVec_\SatI,\PosVec)$ at $\PosVec^{(k)}$ is as follows:
\begin{align}
	\rho(\PosVec_\SatI,\PosVec) \approx\,& \rho(\PosVec_\SatI,\PosVec^{(k)}) + \frac{x^{(k)}-x_\SatI}{\rho(\PosVec_\SatI,\PosVec^{(k)})}\big(x-x^{(k)}\big) \nonumber\\
	&+ \frac{y^{(k)}-y_\SatI}{\rho(\PosVec_\SatI,\PosVec^{(k)})}\big(y-y^{(k)}\big) + \frac{z^{(k)}-z_\SatI}{\rho(\PosVec_\SatI,\PosVec^{(k)})}\big(z-z^{(k)}\big).
\end{align}
Thus, defining $\rho_\SatI^{(k)}\!\triangleq\! \rho(\PosVec_\SatI,\PosVec^{(k)})$, 
$\boldsymbol{\delta}^{(k)}\!\triangleq\!\big[\hat{R}_\SatI-\rho_\SatI^{(k)}\big]_{\SatI\in\setS}\!\in\opR^{|\setS|}$, and the matrix
\begin{align}
	\!\!\bA\!^{(k)} \triangleq
	\begin{bmatrix}
		\frac{x^{(k)}-x_\SatI}{\rho_\SatI^{(k)}} & \frac{y^{(k)}-y_\SatI}{\rho_\SatI^{(k)}} 
		& \frac{z^{(k)}-z_\SatI}{\rho_\SatI^{(k)}} & 1
	\end{bmatrix}_{\SatI\in\setS} \in \opR^{|\setS|\times4},\!\!
\end{align}
the problem \eqref{eq:gnss_pos_opt} can be approximately solved via iterations
\begin{align}
\!\!\!\!\begin{bmatrix}
	\PosVec \\ \cdt
\end{bmatrix}^{\!(k+1)}\!\!\!
&=\argmin_{\PosVec,\cdt} \bigg\| \boldsymbol{\delta}^{(k)} \!+\! \bA\!^{(k)}\!\begin{bmatrix}\PosVec^{(k)} \\ 0~\end{bmatrix} 
	\!-\! \bA\!^{(k)}\!\begin{bmatrix}\PosVec \\ \cdt \end{bmatrix} \bigg\|^2 \!\! \label{eq:gnss_linearized_ls} \\
	&=  \begin{bmatrix}\PosVec^{(k)} \\ 0~\end{bmatrix} + \pinv{\Big(\bA\!^{(k)} \Big)}\boldsymbol{\delta}^{(k)},
\end{align}
where $\PosVec^{(k_{\max})}$ gives the estimate of the receiver coordinates. A typical initialization 
is the Earth's center $\PosVec^{(0)}=\tp{[0,0,0]}$~\cite{esa2013vol1}.

\section{Jammer and Spoofer Mitigation} \label{sec:methods}
Jammers and spoofers interfere with the traditional GNSS receiver from \fref{sec:gnss_basics} in different ways: 
Jammers try to impede the signal acquisition and/or pseudorange estimation entirely, 
while spoofers try to induce erroneous pseudoranges at the receiver based on spuriously acquired signals. 
We therefore mitigate jammers at the signal acquisition stage (\fref{sec:acquisition_defense}), 
whereas spoofed signals are identified and rejected during the pseudorange estimation stage (\fref{sec:doa_defense}). 
Moreover, to increase the robustness of positioning in case that a spoofer manages to bypass the spoofer rejection stage,
we use an outlier-resistant position estimation procedure (\fref{sec:weighted_positioning}).

\subsection{Jammer-Resistant Signal Acquisition} \label{sec:acquisition_defense}

\begin{figure*}[tp]
    \centering
    \subfigure[Baseline acquisition]{
        \includegraphics[width=\caffigwidth]{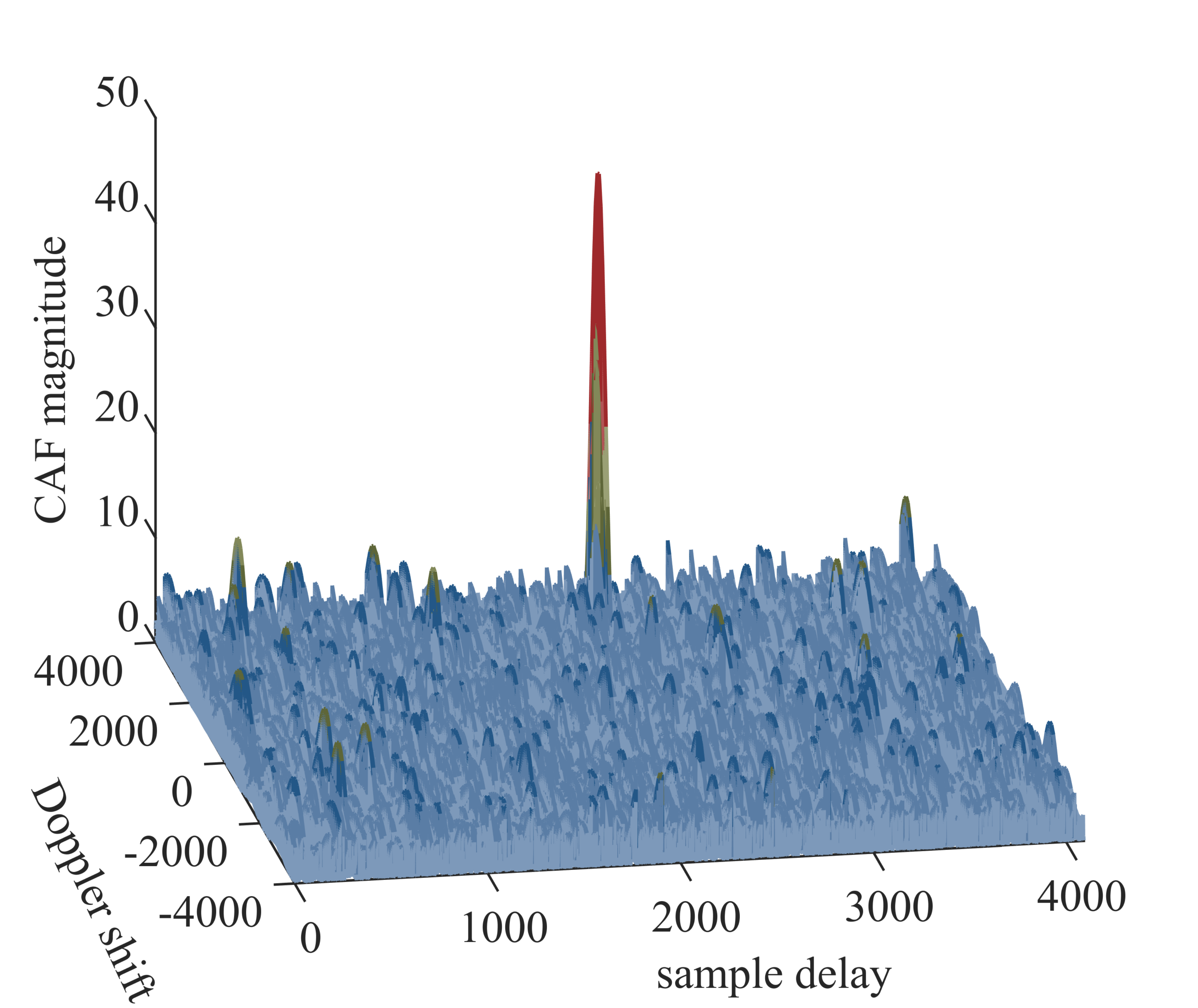}
        \label{fig:caf_baseline_nojam}
    }
    \hspace{0mm}
    \subfigure[Baseline acquisition jammed]{
        \includegraphics[width=\caffigwidth]{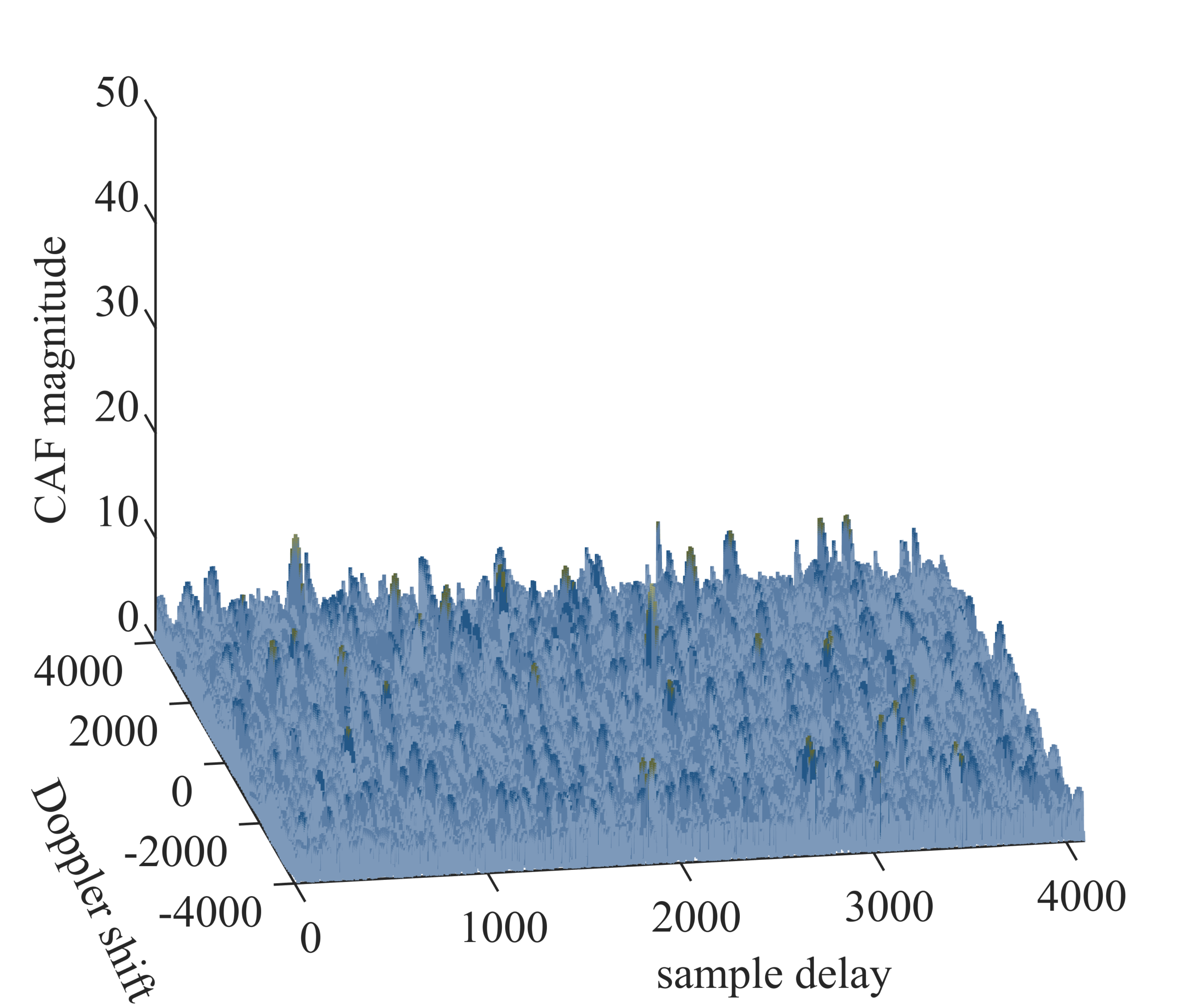}
        \label{fig:caf_baseline_jam}
    }
    \hspace{0mm}
    \subfigure[\gls{jass} acquisition under jamming]{
        \includegraphics[width=\caffigwidth]{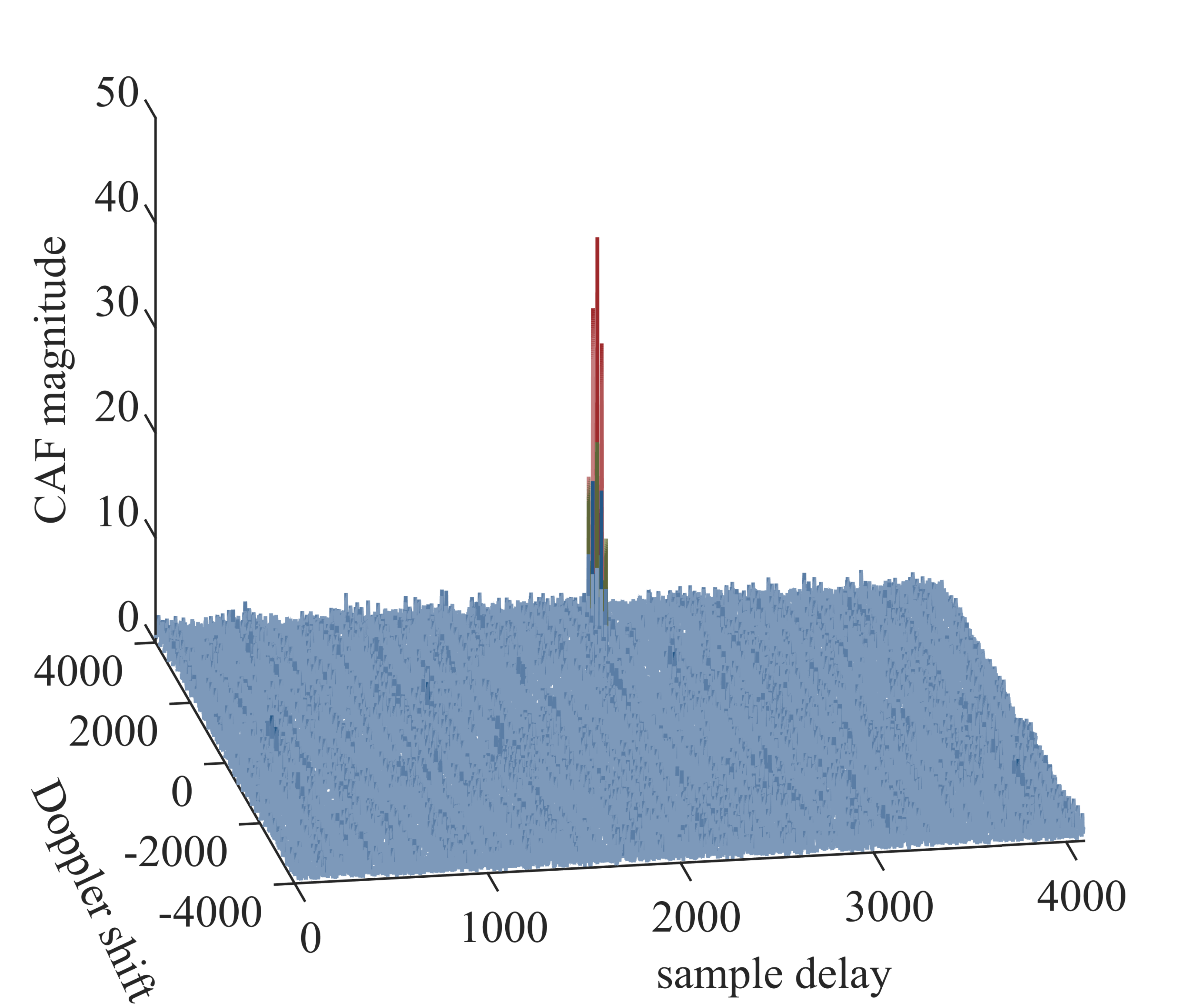}
        \label{fig:caf_jass_jam}
    } 
    \hspace{0mm}
    \subfigure[\gls{jass} acquisition under spoofing]{
        \includegraphics[width=\caffigwidth]{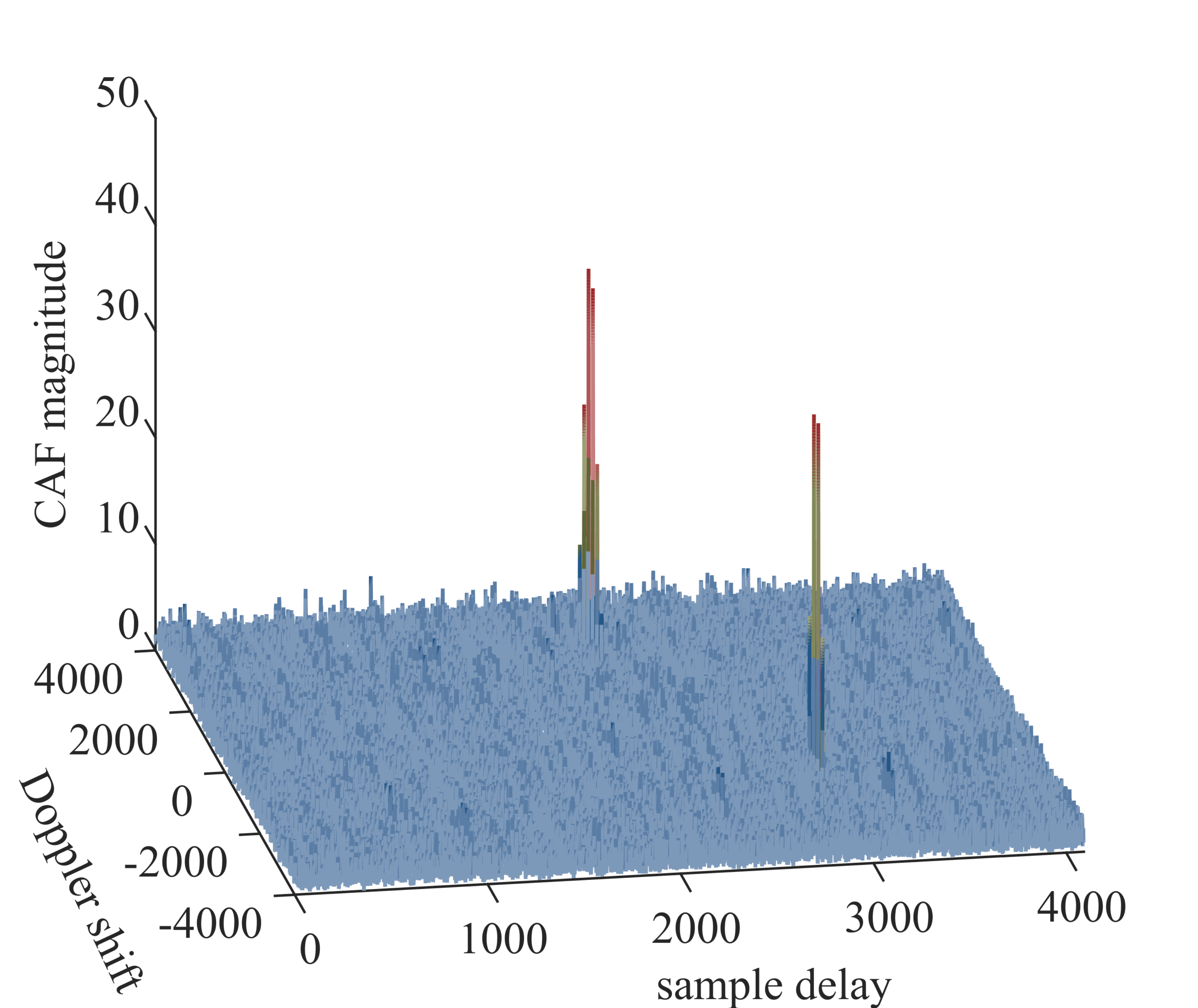}
        \label{fig:caf_jass_spf}
    }
    \caption{Illustration of the traditional \gls{caf} $\SCAF\sb{\SatI}[\JassPhase,f]$ without 
    \fref{fig:caf_baseline_nojam}) and with jamming (\fref{fig:caf_baseline_jam}), as well as of the jammer-resilient 
    \gls{caf} $\widetilde{\SCAF}\sb{\SatI}[\JassPhase,f]$ with jamming (\fref{fig:caf_jass_jam}) and with 
    spoofing (\fref{fig:caf_jass_spf}). 
    The figures are for the L1 C/A \gls{prn} 2 code, sampled at four times the chiprate ($\JassWindowLen=4092$).
    \fref{fig:caf_baseline_nojam} shows a receiver with $B=8$ antennas at $\SNR=-20\;\text{dB}$ 
    (cf. \fref{sec:setup} for definitions of $\SNR$ and $\JSR$) in absence of jamming. 
    In \fref{fig:caf_baseline_jam}, a single-antenna barrage jammer with $\JSR=30\,\text{dB}$ is active and 
    obfuscates the satellite signal, such that the traditional acquisition fails. 
    \fref{fig:caf_jass_jam} shows the same situation, but for the jamming resistant acquisition from \fref{sec:acquisition_defense} (with $\hat{I}=4$), which not only null-steers jamming but also improves 
    the noise floor. 
    \fref{fig:caf_jass_spf} shows that jammer-resistant acquisition remains vulnerable to spoofing, 
    since a spoofer with $\SSR=0\,\text{dB}$ in this example causes a second, equally high, peak in the \glspl{caf}.
    }
    \label{fig:acquisition_overview}
\end{figure*}

The goal of our modified (compared to \fref{sec:acquisition}) signal acquisition is to determine 
the presence, phase, and Doppler shift of satellite signals
in the receive signal even under jammer interference. 
Signals from other satellites appear as (weak) noise after despreading with $\SigCodeVec_\SatI$ 
(see \eqref{eq:other_satellite_interference} and also \fref{fig:caf_baseline_nojam}).
Jamming signals likewise appear as noise; but because of their much larger energy, this noise tends to drown 
 the desired signal (see \fref{fig:caf_jass_jam}).\footnote{Spoofer signals appear as additional peaks in the 
\gls{caf} if the spoofer spoofs the $\SatI$th satellite (see \fref{fig:caf_jass_spf}) and as noise akin to the 
signals from the other satellites otherwise.}
However, what distinguishes the noise due to jammer interference is that it is confined to a rank-$\JamN$ subspace of the receive
signal space \cite{marti2023universal}, so that adaptive spatial filtering can be used to identify and null the interference
subspace. 

To this end, we leverage JASS \cite{marti2025sync}, a recently developed method for jammer-resilient time synchronization, 
with which we enhance the signal acquisition described in \fref{sec:acquisition}: 
Consider the projection matrix
\begin{align}
    \matT\sb{\SatI} &\triangleq \matI\sb{\JassWindowLen}-\conj{\SigCodeVec\sb{\SatI}}\tp{\SigCodeVec\sb{\SatI}}/\norm{\SigCodeVec\sb{\SatI}}\sb{2}^2, \label{eq:code_projection}
\end{align}
which projects (in the time domain) a signal onto the orthogonal complement of the subspace spanned by the spreading code 
$\tp{\SigCodeVec_\SatI}$. 
This property can be leveraged for separating the $\SatI$th satellite's signal of interest from interfering signals
by projecting the windowed and Doppler-compensated (for the candidate Doppler shift $f$) receive signal 
$\matY[\JassPhase]\dg(f)$ (see \eqref{eq:y_window} and \eqref{eq:dopp_diag}) onto said subspace:
\begin{align}
    \widetilde{\matY}\sb{\SatI}(\JassPhase,f) &\triangleq \matY[\JassPhase]\dg(f)\matT\sb{\SatI}. \label{eq:y_window_codeproj}
\end{align}
\emph{If} the sample window is correctly aligned to a signal present in the receive samples using this spreading code 
(i.e., if $\JassPhase=\JassPhase_\SatI$) and if the candidate Doppler compensation matches the true Doppler shift 
(i.e., if $f=f_\SatI$), then the projection $\bT_{\SatI}$ eliminates the $\SatI$th satellite's transmit signal. 
The resulting matrix $\widetilde{\matY}\sb{\SatI}$ therefore consists entirely of interfering signals (from jammers
as well as other satellites) and noise. The dominant $\JassEstInterf$ spatial dimensions\footnote{The number of jammers $\JamN$ 
would presumably be unknown in practice, so $\JassEstInterf$ corresponds to an estimate thereof. Overestimating $\JamN$ is 
not critical, but underestimating it is, see \cite{marti2025sync} and the experiments in \fref{sec:eval}.} 
of this interference can be identified by 
performing an eigenvalue decomposition of the matrix $\widetilde{\matY}\sb{\SatI}(\JassPhase,f)\herm{\widetilde{\matY}\sb{\SatI}(\JassPhase,f)}$,
\begin{align}
    \widetilde{\matY}\sb{\SatI}(\JassPhase,f)\herm{\widetilde{\matY}\sb{\SatI}(\JassPhase,f)}
    &= \begin{bmatrix} \MSubSp{\JassEstInterf} & \MSubSp{n} \end{bmatrix}
    \begin{bmatrix} \dg\sb{\JassEstInterf} & \boldsymbol{0} \\ \boldsymbol{0} & \dg\sb{n}\end{bmatrix} 
    \begin{bmatrix} \herm{\MSubSp{\JassEstInterf}} \\ \herm{\MSubSp{n}} \end{bmatrix},
	\label{eq:eigenvalue_dec}
\end{align}
where $\dg\sb{\JassEstInterf}\in\opR_{\geq0}^{\JassEstInterf\times\JassEstInterf}$ is a diagonal matrix that 
contains the $\JassEstInterf$ largest 
eigenvalues of $\widetilde{\matY}\sb{\SatI}(\JassPhase,f)\herm{\widetilde{\matY}\sb{\SatI}(\JassPhase,f)}$
and $\MSubSp{\JassEstInterf}\in\complexset\sp{\AntN\times\JassEstInterf}$ contains the corresponding orthonormal 
eigenvectors. 
We identify the image of $\MSubSp{\JassEstInterf}$ with the estimated interference subspace, and the image 
of $\MSubSp{n}$ with the estimated noise subspace. The interference subspace can now be nulled using the orthogonal projection
\begin{align} \label{eq:jass_proj}
    \JassProjMat={\matI\sb{\AntN}-\MSubSp{\JassEstInterf}\pinv{\MSubSp{\JassEstInterf}}},
\end{align}
which projects (in the spatial domain) a signal onto the orthogonal complement of the image of $\MSubSp{\JassEstInterf}$. 

Following JASS \cite{marti2025sync}, we now modify the regular \gls{caf} from~\eqref{eq:matrix_caf} to include 
the interference-nulling projection from~\eqref{eq:jass_proj}:
\begin{align} \label{eq:jass_caf}
    \widetilde{\SCAF}\sb{\SatI}[\JassPhase,f] &\triangleq \dfrac{\norm{\JassProjMat\matY[\JassPhase]\dg(f) \conj{\SigCodeVec\sb{\SatI}}}_2^2}{\norm{\JassProjMat\matY[\JassPhase]}_F^2}.
\end{align}
As in the traditional acquisition in \eqref{eq:normal_acquisition}, 
we now try to acquire the $\SatI$th satellite's signal by searching over all possible code phases and Doppler shifts.
A na\"ive way of doing this would be to replace $\SCAF\sb{\SatI}[\JassPhase,f]$ in \eqref{eq:normal_acquisition}
with $\widetilde{\SCAF}\sb{\SatI}[\JassPhase,f]$ from \eqref{eq:jass_caf}. 
However, if we only extract the (single) maximizing argument, we risk that a spoofer which spoofs the $\SatI$th satellite's
signal would create a second \gls{caf} peak that is higher than the legitimate one (cf. \fref{fig:caf_jass_spf}), 
in which case only the spoofed signal would be acquired. 
To circumvent this issue, we acquire \emph{all} signals corresponding to a \gls{caf} peak, which enables us to identify 
and reject spoofed signals later on (see \fref{sec:doa_defense}). To this end, we determine the set of signals satisfying
\begin{align} 
\label{eq:jass_acquisition}
   \setV_\SatI \triangleq \Bigl\{
    (\JassPhase,f) \in\{1,\dots,\JassWindowLen\}\times \setF
    \;\Big|\;    
    \widetilde{\SCAF}\sb{\SatI}[\JassPhase,f] \geq\tau\sb{J} \nonumber \\ 
    \text{ and }   
    \widetilde{\SCAF}\sb{\SatI}[\JassPhase,f] \text{ is a local maximum}
    \Bigr\}.
\end{align}
Note that $|\setV_\SatI|>1$ is a clear indication that the $\SatI$th satellite is being spoofed. 

We define the total set of all acquired signals as
\begin{align}
	\setV \triangleq \bigcup_{\SatI=1}^{S}  \setV_\SatI, 
\end{align}
and we use $\SatI(\nu)$ for $\nu\in\setV$ to denote the index of the satellite which matches the signal $\nu$
(i.e., $\nu\in\setV_{\SatI(\nu)}$ for all $\nu\in\setV$). 
For every acquired signal $\nu\in\setV$, the receiver then estimates the corresponding pseudorange $\hat{R}_{\SatI(\nu)}$ 
as in \eqref{eq:est_pseudorange}.

\subsection{Spoofer-Resistant Pseudorange Estimation} \label{sec:doa_defense}
The goal of our modified (compared to \fref{sec:decode}) pseudorange estimation is to identify 
which of the signals acquired in \fref{sec:acquisition_defense} correspond to spoofers 
in order to reject them and only end up with pseudorange estimates of legitimate satellites. 

To this end, we first estimate the \gls{doa} of all acquired signals. We then develop a novel approach that
tests the consistency of pairs of signals by comparing their respective \gls{doa}s and pseudoranges 
with the almanac \emph{without} any knowledge of the receiver's position. 
Finally, using these pairwise consistency tests, we try to identify the set of legitimate signals.

\subsubsection{Direction of Arrival Estimation} \label{sec:doa_estimation}

For each acquired signal $\nu\in\setV$, let $(\hat\JassPhase_\nu, \hat{f}_\nu)=\nu$ denote the corresponding 
code phase and Doppler estimate (cf. \eqref{eq:jass_acquisition}), 
and let $\JassProjMat\sb{\ValSetI}$ denote the corresponding projection matrix from \eqref{eq:jass_proj}.
We now despread the receive signal as in the traditional GNSS receiver in \fref{sec:decode}, but adding in 
the jammer-mitigating projection~$\JassProjMat\sb{\ValSetI}$. 
Assuming the accuracy of $\hat\JassPhase_\nu$ and $\hat{f}_\nu$, we obtain
\begin{align} \label{eq:symbol_vector_despread_projected}
    \bmr\sb{\ValSetI}[K] 
    &= \JassProjMat\sb{\ValSetI}\matY[K\JassWindowLen +\JassPhase\sb{\ValSetI}]\dg(\hat{f}_\ValSetI) \conj{\SigCodeVec\sb{\SatI(\ValSetI)}} \\
    &= \JassProjMat\sb{\ValSetI}\big(\alpha\sb{\SatI}\,\veca\sb{\SatI}d\sb{\SatI}[K-\Delta K\sb{\SatI}] \norm{\SigCodeVec\sb{\SatI}}^2 
    + \tilde{\vecn}\big),
\end{align} 
where $\tilde\bmn$ subsumes residual disturbance from the other satellite signals and thermal noise, as well as any residual 
jamming interference. 
Estimating the signal's \gls{doa} is tantamount to estimating the steering vector $\veca\sb{\SatI}$; see \eqref{eq:steering_vector}. 
To this end, we use a modified version of the MUSIC algorithm that takes into account the spatial filter
$\JassProjMat\sb{\ValSetI}$'s influence on the receive signal. 
As in the traditional MUSIC algorithm \cite{schmidt1986music}, we collect a window 
$\matR_\nu=\sbr{\bmr\sb{\ValSetI}[K],\ldots,\bmr\sb{\ValSetI}[K+\MACorrN-1]}\in\complexset\sp{\AntN\times\MACorrN}$ of 
receive vectors from which we compute the empirical spatial covariance matrix of $\bmr_\nu[\cdot]$ as 
$\frac{1}{Z}\matR_\nu\herm{\matR_\nu}$. 
We then perform an eigenvalue decomposition of the empirical covariance matrix,
\begin{align}
   \frac{1}{Z}\matR_\nu\herm{\matR_\nu}
    &= \begin{bmatrix} \bme_1 & \MSubSp{N} \end{bmatrix}
    \begin{bmatrix} \sigma_1 & \boldsymbol{0} \\ \boldsymbol{0} & \dg\sb{N}\end{bmatrix} 
    \begin{bmatrix} \herm{\bme_1} \\ \herm{\MSubSp{N}} \end{bmatrix},
\end{align}
where $\sigma_1$ is the principal eigenvalue with corresponding eigenvector $\bme_1$, 
and where $\MSubSp{N}\in\opC^{B\times(B-1)}$ contains the eigenvectors corresponding to the remaining eigenvalues.

To estimate the \gls{doa}, we consider a grid of testing angles $\theta\in\Theta\subset[0,\pi/2]$ and $\varphi\in\Phi\subset[-\pi,\pi]$
with corresponding steering vectors $\bma(\theta,\varphi)$ as in \eqref{eq:steering_vector}. 
To take into account the fact that the spatial filter $\JassProjMat\sb{\ValSetI}$ nulls signals from certain directions, 
we estimate the \gls{doa} of the signal $\nu\in\setV$ as follows: 
\begin{align}
    (\hat\theta_\nu, \hat\varphi_\nu) &= \argmin_{(\theta,\varphi)\in\Theta\times\Phi}
	\norm{\herm{\MSubSp{N}}\dfrac{\JassProjMat\sb{\ValSetI}\MSteerVecEst(\theta,\varphi)}{\|\JassProjMat\sb{\ValSetI}\MSteerVecEst(\theta,\varphi)\|}}^2 \\
	&= \argmax_{(\theta,\varphi)\in\Theta\times\Phi}
	\frac{\|\JassProjMat\sb{\ValSetI}\MSteerVecEst(\theta,\varphi)\|^2}{\|\herm{\MSubSp{N}}\JassProjMat\sb{\ValSetI}\MSteerVecEst(\theta,\varphi)\|^2}, \label{eq:music_modified}
\end{align}
and we define the objective in \eqref{eq:music_modified} to be zero if 
$\JassProjMat\sb{\ValSetI}\MSteerVecEst(\theta,\varphi)=\boldsymbol{0}$. 
The corresponding steering vector is denoted $\hat\bma_\nu\triangleq\bma(\hat\theta_\nu, \hat\varphi_\nu)$. 

This \gls{doa} estimation approach builds upon the idea that the acquired signal $\nu$ impinges under LoS conditions, 
which we assumed was the case for unobstructed satellite signals. However, we assumed 
that spoofed signals could impinge under Rayleigh fading conditions as well as LoS conditions (cf. \fref{sec:signal_model}). 
If the objective value of the maximizing argument of \eqref{eq:music_modified} is small, 
then it indicates that the impinging signal does not exhibit LoS conditions, which---given our assumptions---can be
used as a preliminary indicator for identifying spoofed signals. Hence, we reject all signals $\nu\in\setV$ 
for which the maximal objective value in \eqref{eq:music_modified} is smaller than some threshold~$\MTh$, 
and we denote the remaining set of signals by~$\breve\setV$.

\subsubsection{Pairwise Consistency Test} \label{sec:plausibility}
%
We now introduce a novel approach to test the consistency of pairs of signals by comparing their respective \gls{doa}s 
and pseudoranges with the almanac \emph{without} knowing the receiver's position.
Assume for the moment that both $\nu$ and $\nu'\in\breve\setV$ correspond to legitimate signals from two different satellites
$\SatI(\nu)$ and $\SatI(\nu')$, respectively. 
We therefore consider the difference between the known (from the almanac) positions $\SPs$ and $\SPr$ 
of the satellites $\SatI(\nu)$ and $\SatI(\nu')$, which can be written as
\begin{align}
    \SPs-\SPr &= \Prx \!+\! \rhos\RotMatAnt\vs-\big(\Prx \!+\! \rhor\RotMatAnt\vr\big) \label{eq:pos_difference} \\
    &= \RotMatAnt\del{\rhos\vs-\rhor\vr},
\end{align}
where $\Prx$ is the receiver's (as of yet) unknown position in \gls{ecef} coordinates, 
$\RotMatAnt$ is the unknown rotation matrix which rotates the receiver-centered antenna coordinate system of 
the antenna arrangement (\fref{fig:ray_onto_antennas}) into the \gls{ecef} coordinate system at the receiver position, 
and $\vs\triangleq\bmv\big(\theta_{\varsigma(\nu)}, \varphi_{\varsigma(\nu)}\big)$ and 
$\vr\triangleq\bmv\big(\theta_{\varsigma(\nu')}, \varphi_{\varsigma(\nu')}\big)$ are the unitary direction vectors (cf. the 
right-hand-side in \eqref{eq:steering_vector}) towards the satellites $\varsigma(\nu)$ and $\varsigma(\nu')$, respectively. 
We now take the squared norm of \eqref{eq:pos_difference} and obtain
\begin{align}
    &\big\|\SPs-\SPr\big\|\sp{2} \nonumber \\
    &= \|\rhos\vs-\rhor\vr\|\sp{2} \\
    &= \rhos\sp{2}\|\vs\|\sp{2}-2\rhos\rhor\tp{\vs}\vr+\rhor\sp{2}\|\vr\|\sp{2} \label{eq:SatPosD1}\\
    &= \big(\rhos-\rhor\big)\sp{2} + 2\rhos\rhor -2\rhos\rhor\tp{\vs}\vr, \label{eq:SatPosD2}
\end{align}
where \eqref{eq:SatPosD2} follows from the fact that $\vs$ and $\vr$ are unit vectors, and by completing the square. 
Following the pseudorange relation from \eqref{eq:pseudorange_exact},
we can now rewrite the range difference $\rhos-\rhor$ in \eqref{eq:SatPosD2} as follows:
\begin{align}
	\rhos-\rhor &= \rhos+\cdt - \rhor-\cdt \\
	&= \Rs-\Rr \\
	&\triangleq \Rsr. \label{eq:pseudorange_difference}
\end{align}
Rearranging \eqref{eq:pseudorange_difference} into $\rhor=\rhos-\Rsr$ and 
plugging this into \eqref{eq:SatPosD2} yields the quadratic equation
\begin{align} \label{eq:plausibility_quadratic}
    \|\SPs\!-\!\SPr\|^{2} =& \,\Rsr\sp{2} + 2\rhos(\rhos\!-\!\Rsr) (1\!-\!\tp{\vs}\vr).
\end{align}
Note that the receiver can compute the value of the left-hand-side from the almanac, 
that it can obtain an estimate of $\Rsr$ by computing $\hatRsr=\hatRs-\hatRr$ (where $\hatRs$ and $\hatRr$ 
have been estimated in \fref{sec:acquisition_defense}),
and that it has obtained estimates of $\vs$ and $\vr$ in \fref{sec:doa_estimation}.
Hence, by plugging in these estimates, the receiver can obtain an estimate for the unknown range $\rhos$ using 
the quadratic formula\footnote{The subscript $\nu\leftarrow\nu'$ indicates that the range of the satellite 
correspondig to signal $\nu$ was estimated using the signal $\nu'$ as a reference}
\begin{align}
    \DistEst{\nu}{\nu'} &= \dfrac{-b+\sqrt{b^2-4ac}}{2a}, \label{eq:dist_est_pldoa}
\end{align}
where
\begin{align}
    a &= 2(1-\tp{\hatvs}\hatvr) \\
    b &= -2\hatRsr(1-\tp{\hatvs}\hatvr) \\
    c &= \hatRsr^{2}-\|\SPs-\SPr\|^{2}. \label{eq:crit_c}
\end{align}
Remember that the derivation of \eqref{eq:dist_est_pldoa} (from \eqref{eq:pos_difference} on) was based on the 
premise that both $\nu$ and $\nu'$ correspond to legimitate signals from satellites $\SatI(\nu)$ and $\SatI(\nu')$. 
If this premise is violated,~\eqref{eq:dist_est_pldoa} tends to give bogus results 
because the direction vectors $\hatvs$ and $\hatvr$ are inconsistent with the almanac positions $\SPs$ and $\SPr$ in 
\eqref{eq:pos_difference} (when the spoofer does not successfully replicate the satellite geometry),
or because the difference of estimated pseudoranges $\hatRsr$ is inconsistent with the range difference 
in \eqref{eq:pseudorange_difference} (when the spoofer does not successfully replicate the satellite's timing at the receiver). 

We now use this behavior to our advantage: Given the GNSS satellite orbits, for any location $\PosVec$ on Earth and for 
any satellite~$\SatI$, the range $\rho_\SatI=\|\PosVec_\SatI-\PosVec\|_2$ should be within some range $[\rho_\text{min}, \rho_\text{max}]$. 
Moreover, from the triangle inequality, it follows that $c$ in \eqref{eq:crit_c} should be negative. 
If one or both of these conditions are violated, this is indicative that either $\nu$ or $\nu'$ may not correspond 
to a legitimate satellite signal. For any two signals $\nu$ and $\nu'$, we therefore define the following plausibility function:
\begin{align}
	p(\nu,\nu') = \begin{cases} \label{eq:plausibility_function}
		1 & \text{ if } \DistEst{\nu}{\nu'} \in [\rho_\text{min}, \rho_\text{max}] \text{ and } c<0 \\
		0 & \text{ else}.
	\end{cases}	
\end{align}
Using this function, we can generate the plausibility matrix $\bG\!\in\!\{0,1\}^{|\setV|\times|\setV|}$
whose entries are given by \mbox{$p(\nu,\nu'),\,\nu,\nu'\!\in\!\setV$.}
Note that $p(\nu,\nu')\neq p(\nu',\nu)$ because $\DistEst{\nu}{\nu'}\neq \DistEst{\nu'}{\nu}$, 
and so~$\bG$ is not symmetric. However, we can get a symmetric plausibility matrix $\bG_{\text{sym}}$ 
from $\bG$ as
\begin{align}
    \bG_{\text{sym}} = \bG\had\tp{\bG}.
\end{align}
This matrix can be interpreted as the adjacency matrix of an undirected graph $\setG=(\setV,\setE)$
whose vertices are the acquired signals $\setV$. Two signals $\nu$ and $\nu'$ are connected within $\setG$ 
if and only if $p(\nu,\nu')p(\nu',\nu)=1$, i.e., if the relative range estimates $\DistEst{\nu}{\nu'}$ and $\DistEst{\nu'}{\nu}$
are both plausible according to \eqref{eq:plausibility_function}. 

\subsubsection{Identifying the Set of Legitimate Signals}
A complete subgraph of $\setG$ corresponds to a set of signals are all consistent with each other in the 
sense of satisfying the plausibility function in \eqref{eq:plausibility_function}. 
We therefore try to infer the set of legitimate signals by searching for the largest complete subgraph of $\setG$, 
i.e., for the maximum clique in $\setG$ \cite{Karp1972CliqueNP}. 
However, finding the maximum clique of a graph is an NP-hard problem \cite{Karp1972CliqueNP}, 
so we use an approximate method based on a greedy approach.

\begin{figure}[tp]
    \centering
	\!
    \subfigure[\!Iteration $t\!=\!1$]{
        \includegraphics[width=0.21\columnwidth]{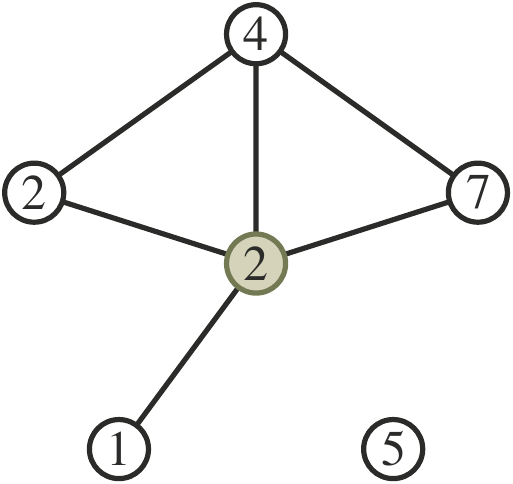}
        \label{fig:example_a}
    }
    \hspace{-1.5mm}
    \subfigure[\!Iteration $t\!=\!2$]{
        \includegraphics[width=0.21\columnwidth]{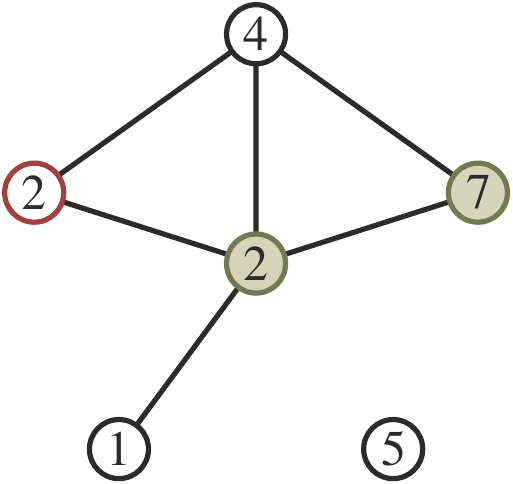}
        \label{fig:example_b}
    }
    \hspace{-1.5mm}
    \subfigure[\!Iteration $t\!=\!3$]{
        \includegraphics[width=0.21\columnwidth]{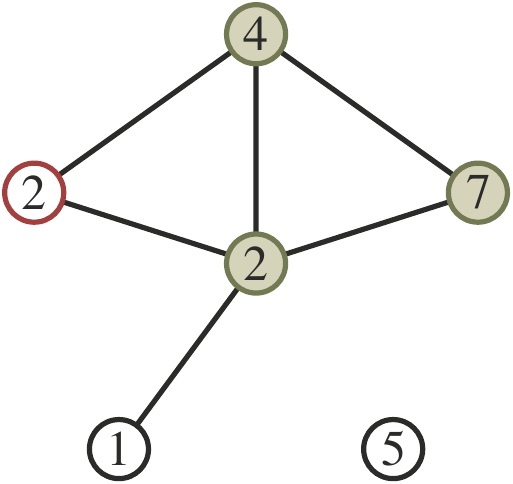}
        \label{fig:example_c}
    } 
    \hspace{-1.5mm}
    \subfigure[\!Final iterations]{
        \includegraphics[width=0.21\columnwidth]{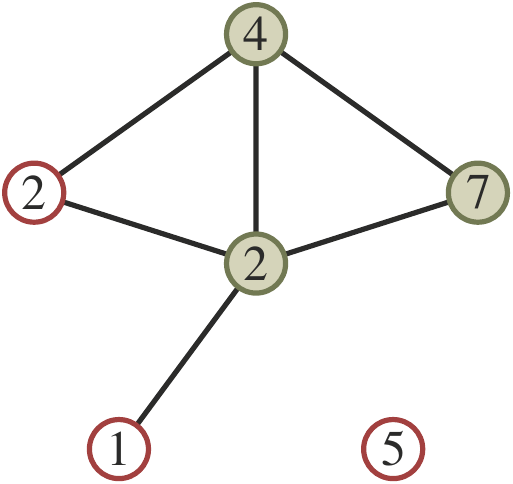}
        \label{fig:example_d}
    }
	\!
	\caption{Illustration of the greedy clique finding algorithm of \eqref{eq:clique_greedy}. In the first iteration, 
	the vertex of maximum degree is added to the clique (shown in green), whose satellite index is $\SatI=2$. 
	In the second iteration, another signal with satellite index $\SatI=2$ is passed over (shown in red), 
	and the tie between the $\SatI=4$ and $\SatI=7$ signals is arbitrarily broken in favor of the latter. 
	In the third iteration, the $\SatI=4$ signal is added to the clique. In the remaining iterations, 
	there are no vertices that can be added to the clique and the algorithm terminates accordingly.}
    \label{fig:greedy_graph}
\end{figure}

We first search for the vertex with the highest degree, with which we initialize the clique:
$\text{clique}^{(0)}=\argmax_{\nu\in\setV}\text{deg}(\nu)$.\footnote{If there are multiple maximizing vertices, 
we select one of them at random. The same holds for \eqref{eq:clique_greedy}.}
We then iteratively augment the clique with the vertex that has the highest degree among all remaining vertices which
(i) is connected with all nodes that are already included in the clique, 
and which (ii) does not correspond to a satellite that is already represented in the clique (since each satellite 
can have at most one legitimate signal):
\begin{align}
	\text{clique}^{(t+1)} = \text{clique}^{(t)} \cup \argmax_{\substack{\nu\in\setV\\ \forall\nu'\in \text{clique}^{(t)}: (\nu,\nu')\in\setE\\ \forall\nu'\in \text{clique}^{(t)}: \SatI(\nu)\neq\SatI(\nu')}} \text{deg}(\nu). \label{eq:clique_greedy}
\end{align}
We denote the set of signals included in the final clique (i.e., the signals deemed to be legitimate) 
as $\breve{\breve\setV}$. 
An explanatory illustration of this algorithm is provided in \fref{fig:greedy_graph}. The numbers in the vertices designate the satellite ID in the tuple and the operation is illustrated from left to right. 

\subsection{Outlier-Resistant Positioning} \label{sec:weighted_positioning}
The plausibility filtering yields a clique of tuples which are spatially consistent with one another within the constellation's dimensional relations.
Spoofed signals that closely match the \gls{doa} of the corresponding legitimate signal but that are marginally shifted 
in terms of the pseudorange can potentially bypass the spoofer rejection. 
For this reason, we modify the position estimation method to make it resistant to outliers. 

The set of selected signals $\breve{\breve\setV}$ corresponds to the pseudorange estimates $\hat{R}_{\SatI(\nu)}, \nu\in\breve{\breve\setV}$ 
and satellite positions $\PosVec_{\SatI(\nu)}, \nu\in\breve{\breve\setV}$ required to estimate the position of the receiver.
Instead of iteratively solving the linearized least-squares problems \eqref{eq:gnss_linearized_ls}, 
which give each acquired pseudorange $\SatI$ the same weight, we use an \gls{irls} approach \cite{Gentle2007matrixIRLS} 
in which the different pseudorange estimates $\hat{R}_{\SatI(\nu)}, \nu\in\breve{\breve\setV}$ are weighted with 
an iteration-dependent weighting matrix $\bW^{(k)}=\text{diag}(w_1^{(k)},\dots,w_{|\breve{\breve\setS}|}^{(k)})$. 
That is, instead of iterating \eqref{eq:gnss_linearized_ls}, we iterate
\begin{align}
&\begin{bmatrix}
	\PosVec \\ \cdt
\end{bmatrix}^{\!(k+1)} \nonumber \\
	&=\argmin_{\PosVec,\cdt} \bigg\| \Big(\bW^{(k)}\Big)^{\!\frac12}\bigg(\boldsymbol{\delta}^{(k)} \!+\! \bA\!^{(k)}\!\begin{bmatrix}\PosVec^{(k)} \\ 0~\end{bmatrix} 
	\!-\! \bA\!^{(k)}\!\begin{bmatrix}\PosVec \\ \cdt \end{bmatrix}\bigg) \bigg\|^2 \!\! \\
	&=  \begin{bmatrix}\PosVec^{(k)} \\ 0~\end{bmatrix} + 
	\inv{\big(\tp{\big(\bA\!^{(k)\!}\big)}\bW^{(k)}\bA\!^{(k)}\big)}\tp{\big(\bA\!^{(k)}\big)\!}\bW^{(k)} \boldsymbol{\delta}^{(k)},
\end{align}
where $\boldsymbol{\delta}^{(k)}\triangleq\big[\hat{R}_{\SatI(\nu)}-\rho_{\SatI(\nu)}^{(k)}\big]_{\nu\in\breve{\breve\setV}}\in\opR^{|\breve{\breve\setV}|}$ 
with $\rho_{\SatI(\nu)}^{(k)}\triangleq \rho(\PosVec_{\SatI(\nu)},\PosVec^{(k)})$, 
and where diagonal entries $w_{\SatI(\nu)}^{(k)}>0$ of the weighting matrix  $\bW^{(k)}$ correspond to the reliabilities 
of the respective acquired pseudoranges. These reliabilities are iteratively computed from the data as
\begin{align}
	w_\SatI^{(k+1)} = \bigg(\frac{1}{\max\{\sigma,e_{\SatI(\nu)}^{(k+1)}\}}\bigg)^2,
\end{align}
where
\begin{align}
	\!\!\!&\bme^{(k+1)} \nonumber \\
	\!\!&\!\!= \boldsymbol{\delta}^{(k)} + \bA\!^{(k)}\!\begin{bmatrix}\PosVec^{(k)} \\ 0~\end{bmatrix}	
	- \bA\!^{(k)}\!\begin{bmatrix}\PosVec \\ \cdt \end{bmatrix}^{\!(k+1)} \\
	\!\!&\!\!= \Big(\bI_{|\breve{\breve\setV}|} \!-\! \bA\!^{(k)}\inv{\big(\tp{\big(\bA\!^{(k)\!}\big)}\bW^{(k)}\bA\!^{(k)}\big)}\!\tp{\big(\bA\!^{(k)}\big)\!}\bW^{(k)}\Big) \boldsymbol{\delta}^{(k)}\!\!,\!\!
\end{align}
and where $\sigma=\frac{1}{6}cT$ represents the intrinsic imprecision of the acquired pseudoranges.\footnote{
This value for $\sigma$ is obtained by assuming that the pseudorange measurement error is uniformly distributed
on the interval $[-0.5\,\SoL T,+0.5\,\SoL T]$, and thus has standard deviation proportional to $\SoL T$. 
The factor $1/6$ is chosen empirically.
}

\section{Evaluation in Simulation} \label{sec:eval}

\subsection{Simulation Setup} \label{sec:setup}
We now evaluate the efficacy of SCHIEBER using MATLAB simulations. Our simulator models the physical constellation of \gls{gps} satellites, a receiver positioned on Earth, and the \gls{gps} L1 C/A signals arriving onto its antenna arrangement. 
Within this framework, the influence of different attack scenarios on positioning is assessed.
The satellite positions are initialized at the reference epoch positions \cite{GPS_SPSPS}, from where we simulate their trajectories along their ideal trajectories up to the timestamp of simulation. 
For simplicity of the simulation, we assume circular orbits without perturbations.
We assume that the receiver is positioned on the surface of the Earth, which we model as a perfect sphere. 
For every satellite in the \gls{gps} constellation, 
we deduce the satellite visibility (or lack thereof) above the receiver's horizon.
For every visible satellite, we compute the distance, velocity towards the receiver, 
and the \glspl{doa} of the incoming signal at the receiver. 
From these values the signal properties for each receivable satellite follow from our model in \fref{sec:signal_model}, 
which yields the satellite signal part of \eqref{eq:io_model}.

We model the satellite signal attenuation using a simple distance-based model: the mean distance $\rho_0$ of a \gls{los} satellite ($\rho_0\approx23'000\;$km) is normalized such that it corresponds to a channel gain of $\|\vech\sb{0}\|^2=\AntN$ after channel attenuation as in \eqref{eq:los_channel}. All other distance-induced attenuations are scaled according to their distance with respect to this mean distance $\rho_0$ using the free-space path loss model 
\begin{align}
    \|\vech\sb{\SatI}\| = \|\vech_0\|\frac{\rho_0}{\rho\sb{\SatI}}.
\end{align}
The noise power is specified in terms of the mean received \gls{snr} using the mean satellite path gain~$\|\vech\sb{0}\|^2$: 
\begin{align}
    \SNR=\dfrac{\Exop\sbr{\|\vech\sb{0}t_\SatI[\cdot]\|^2}}{\Exop\sbr{\|\vecn[k]\|^2}}
    = \dfrac{B}{\Exop\sbr{\|\vecn[k]\|^2}}.
\end{align} 
Both for \gls{schieber} and for the performance baseline, the receiver is equipped with $\AntN=8$ antennas that are arranged in a ring, 
with neighboring antennas being separated by half a wavelength. In the simulations, the receiver remains stationary such that the satellite Doppler frequency range is $\pm4\;$kHz. The signals are sampled at four times their chiprate, i.e., 
at $4.092\;$MHz.\footnote{
Reference \cite{akospini2006sampfreqGNSS} discourages whole integer oversampling due to precision loss caused by the chip duration uncertainty. Our design choice is motivated by simplified front-end simulation at the cost of higher positioning errors, which would allow future works to improve this aspect.} 
The acquisition is performed over a single spreading code of length $\JassWindowLen=4092$ and over a Doppler granularity of $\delta f=250\;$Hz. We empirically set the acquisition threshold for the baseline and for \gls{schieber} to $\tau=11.2$ and $\tau_J=7.5$, respectively. 
The jammer-resistant acquisition rejects an interference subspace of dimension $\JassEstInterf=4$.
For the \gls{doa} estimation, we use a window of $\MACorrN=10$ consecutive vectors, over which the \gls{doa} of all impinging LoS signals is assumed to remain constant. 
The \gls{doa} search is performed over the sets $\Theta=\{0^\circ,1^\circ,\ldots,90^\circ\}$ and 
$\Phi=\{0^\circ,1^\circ,\ldots,359^\circ\}$, and we empirically set the \gls{los} threshold to $\MTh=10$.
The plausibility range for the distance estimation in \eqref{eq:dist_est_pldoa} is set to 
$[\rho_{\min},\rho_{\max}]=[18\,000\text{km}, 28\,000\text{km}]$, which allows for some margin of error (the real distance between an Earth-surface receiver and a GPS satellite above the horizon is in the range of $20'000\;$km to $27'000\;$km).

As performance metric, we consider the empirical \gls{cdf} of the Euclidean distance between the actual position 
and the projection of the estimated position onto the Earth's surface. Each empirical \gls{cdf} is computed from 
200 Monte--Carlo trials. For every trial, the satellites are placed at the reference epoch positions \cite{GPS_SPSPS}
for a randomly drawn time within the year 2024; the receiver is placed uniformly at random on the Earth's surface, 
and jammers as well as spoofers are placed randomly anywhere in the hemisphere above the receiver's horizon. 
Unless noted otherwise, attack signals are assumed to impinge onto the receive antenna array under pure \gls{los} 
conditions. When noted otherwise, we assume i.i.d. Rayleigh fading, i.e., we draw the attacker channels 
$\JamChVec\sb{\JamI}$ and $\SpfChVec\sb{\SpfI}$ from $\mathcal{CN}(\veczero,\matI\sb{\AntN})$.

\subsection{Interference Signals} \label{sec:attacker}
Single and multi-antenna jammers operate as described in \fref{sec:interference_model}, 
so the jammer transmit signals are
\begin{align} \label{eq:sim_jamming}
    \JamSymVec[k] = \begin{cases}
        \JamSym[k]\sim\mathcal{CN}(0,\sigma\sb{J}^2) & \JamN=1 \\
        \matB[k]\vecb[k], \:\vecb[k]\sim\mathcal{CN}(\veczero,\sigma\sb{J}^2\matI_R) & \JamN\geq2, 
    \end{cases}
\end{align}
where the beamforming matrix $\matB[k]$ is defined as described in~\eqref{eq:jammer_multi_ant}, 
and where $R$ is set to $R=\AntN-1$ 
(i.e., for multi-antenna jamming, a randomly chosen jamming antenna is inactive at any point in time).
A jammer's interference power is characterized relative to the mean received satellite signal power in terms of 
the per-transmit-antenna \gls{jsr}
\begin{align}
    \JSR = \dfrac{\Exop\sbr{\|\JamChVec\sb{\JamI}\JamSym\sb{\JamI}[k]\|_2^2}}{\Exop\sbr{\|\vech\sb{0}\SigRX[k]\|_2^2}} .
\end{align}

In our simulation, spoofers only try to impersonate satellites that are visible to the receiver. 
Spoofers start replicating true satellite signals \emph{before} the receiver starts with signal acquisition,
i.e., like satellites, spoofers are active for the entires simulation duration. 
They imitate complete satellite messages even though the actual content is irrelevant in the simulations as long as the pseudorange is measurable. 
The spoofing power is characterized relative to the mean received satellite signal power in form of the per-transmit-signal
(cf. \eqref{eq:spoofer_tx_symbol}) \gls{ssr}
\begin{align}
    \SSR = \dfrac{\frac{1}{\SpfSatN}\Exop\sbr{\|\SpfChVec\sb{\SpfI}\SpfSym\sb{\SpfI}[k]\|_2^2}}{\Exop\sbr{\|\vech\sb{0}\SigRX[k]\|_2^2}}.
\end{align}
In all spoofer scenarios, each spoofer antenna imitates the same number $\SpfSatN$  of satellites. 
We set the delay of the spoofed signals such that they appear at somewhat realistic distance to the receiver in comparison to the legitimate signals. These distances are randomly selected in the range of $20\,000$\,km to $29\,000\,$km for each imitated signal. The Doppler frequency of each spoofed satellite signal is randomly selected in the range of $\pm4\;$kHz, not necessarily mirroring the Doppler shift of the imitated satellite since \gls{schieber} makes no plausibility interpretation of the estimated Doppler shift.

\subsection{Results}

\subsubsection{Performance in Jammer-Free, Spoofer-Free GPS}
\begin{figure}[tp]
    \centering
    \includegraphics[width=\figwidth]{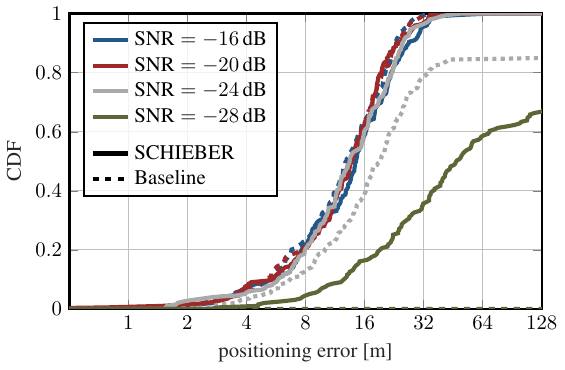}
    \caption{Cumulative distribution function (CDF) of the positioning error of SCHIEBER and the corresponding baseline in the absence of jamming or spoofing.}
    \label{fig:no_interference}
\end{figure}
We start by comparing the positioning performance of \gls{schieber} and the baseline in the absence of jamming and spoofing.
\fref{fig:no_interference} shows empirical CDFs of the positioning error for different SNRs. 
For an SNR of $-20$\,dB or higher, \gls{schieber} and the baseline have virtually identical performance, 
with a positioning error of $30$\,m or less.\footnote{Note that the positioning error of commercial GPS receivers in the L1
band is typically around $2.5\;$m \cite{ubloxNeoM8}. This performance difference is explained by our
simplified receiver architecture, which uses integer oversampling \cite{akospini2006sampfreqGNSS} and does not perform
smoothing of the estimated position over time. Improving both of these aspects is left for future work.} At lower SNRs, \gls{schieber} outperforms the baseline even in the absence
of spoofing or jamming. The reason for this is due to the fact that jammer-resistant signal acquisition stage 
of \gls{schieber} (cf. \fref{sec:acquisition_defense}) also helps to mitigate the mutual interference of the GPS satellites. 
Note that $-24$\,dB and $-28$\,dB are the respective SNRs at which the baseline and \gls{schieber} occasionally fail
to acquire enough satellite signals for successful positioning, which can lead to positioning errors in the 
order of tens of thousands of kilometers.

\subsubsection{Performance Against Jamming}
\begin{figure}[tp]
    \centering
    \includegraphics[width=\figwidth]{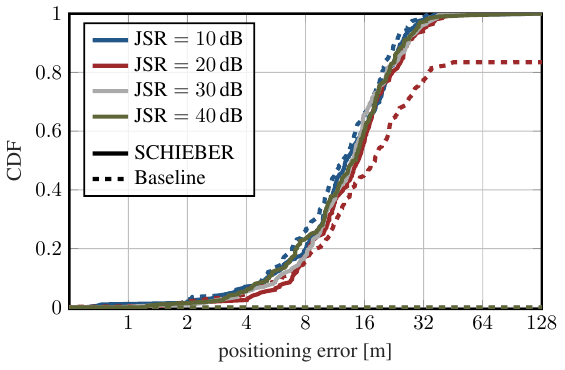}
    \caption{Cumulative distribution function (CDF) of the positioning error of SCHIEBER and the corresponding baseline in the under jamming by a single-antenna jammer, for different jamming-to-signal ratios (JSRs).}
    \label{fig:single_antenna_jammer}
\end{figure}
\begin{figure}[tp]
    \centering
    \includegraphics[width=\figwidth]{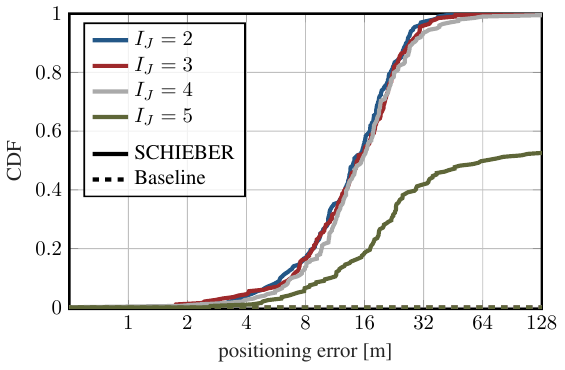}
    \caption{Cumulative distribution function (CDF) of the positioning error of SCHIEBER and the corresponding baseline in the under jamming by a multi-antenna jammer, for different numbers of jammer antennas.}
    \label{fig:multi_antenna_jammer}
\end{figure}
Next, we consider the performance against jammers. 
\fref{fig:single_antenna_jammer} shows the performance against one single-antenna jammer with different 
jamming powers at an SNR of $-20$\,dB. At a \gls{jsr} or $10$\,dB, the received jammer power is negligible
comparable to the noise power, so the baseline achieves comparable positioning performance as in the 
interference-free case (cf. \fref{fig:no_interference}). However, at a \gls{jsr} of $20$\,dB,
the jammer interference becomes noticeable, and the baseline fails to achieve positioning in around $20\%$ of cases. 
For \gls{jsr}s of $30$\,dB or more, the jammer interference becomes critical, and the baseline fails completely. 
However, thanks to the jammer-mitigating signal acquisition (cf. \fref{sec:acquisition_defense}), 
\gls{schieber} achieves comparable performance as in the interference-free case regardless of the jammer power. 
In particular, \gls{schieber} achieves a positioning error of around $30$\,m or less in all cases. 

\fref{fig:multi_antenna_jammer} shows the performance against multiple jammers (ranging 
from $\JamN=2$ to $\JamN=5$) at a \gls{jsr} of $30$\,dB and an \gls{snr} of $-20$\,dB. 
At these jammer powers, the baseline fails completely regardless of the number of jammers.
In contrast, \gls{schieber}, whose jammer-mitigating signal acquisition nulls an interference-subspace of
dimension $\JassEstInterf=4$ (cf. \fref{sec:setup}), successfully mitigates the jammers as long 
as $\JamN\leq\JassEstInterf=4$. In these cases, the positioning performance matches the performance 
of the interference-free case (cf. \fref{fig:no_interference}). 
However, $\JamN=5$ exceeds the dimension $\JassEstInterf$ of the nulled interference-subspace, 
so that \gls{schieber} mitigates the interference only partially, and positioning fails in around $50\%$ of the cases.

\subsubsection{Spoofing}
\begin{figure}[tp]
    \centering
    \includegraphics[width=\figwidth]{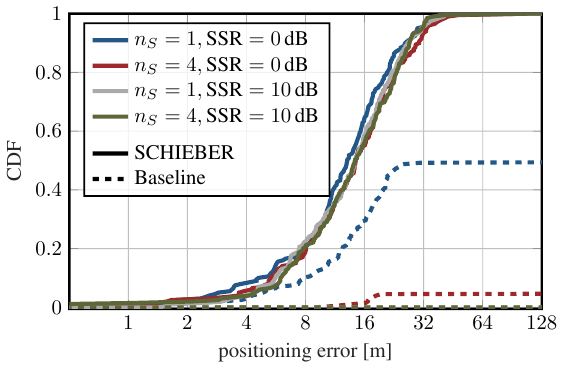}
    \caption{Cumulative distribution function (CDF) of the positioning error of SCHIEBER and the corresponding baseline in the under spoofing by a single spoofer, for different spoofer-to-signal ratios (SSRs) and different numbers of spoofed satellites $\SpfSatN$.}
    \label{fig:single_antenna_spoofer}
\end{figure}
\begin{figure}[tp]
    \centering
    \includegraphics[width=\figwidth]{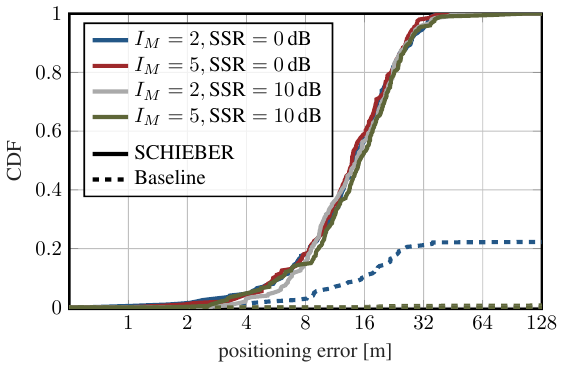}
    \caption{Cumulative distribution function (CDF) of the positioning error of SCHIEBER and the corresponding baseline in the under spoofing by a multiple spoofers, for different spoofer-to-signal ratios (SSRs) and different numbers of spoofers $\SpfN$.}
    \label{fig:multi_antenna_spoofer}
\end{figure}
We now turn to analyzing the performance against spoofers. 
\fref{fig:single_antenna_spoofer} shows the performance (at $\text{SNR}=-20$\,dB) against one single-antenna spoofer 
that simultaneously spoofs the signal of $\SpfSatN=1$ or $\SpfSatN=4$ satellites at $\text{JSR}=0$\,dB or 
$\text{JSR}=10$\,dB. If the spoofer only spoofs a single satellite at $\text{JSR}=0$\,dB, 
it is based purely on luck whether the baseline receiver acquires the legitimate signal or the spoofed signal.
Hence, the baseline fails in $50\%$ of the cases. However, if the spoofer power increases to $\text{JSR}=10$\,dB, 
the baseline receiver consistently acquires the spoofed signal and thus fails completely. 
If the spoofer spoofs $\SpfSatN=4$ satellites at $\text{SSR}=0$\,dB, there is a roughly $1-0.5^4=0.9375$ probability that 
the baseline receiver acquires at least one spoofed signal instead of the corresponding legitimate signal. 
Hence, the baseline achieves successful positioning in less than $10$\% of the cases. 
And if the spoofer power increases to $\text{SSR}=10$\,dB, the baseline again fails completely. 
In contrast, thanks to the spoofer-resistant pseudorange estimation (cf. \fref{sec:doa_defense}), 
\gls{schieber} is able to reject the spoofed signals and achieve positioning with comparable performance as 
in the unspoofed~case. 

\fref{fig:multi_antenna_spoofer} shows the performance (at $\text{SNR}=-20$\,dB) against multiple spoofers, 
where each spoofer spoofs the signal of one satellite, either at $\text{SSR}=0$\,dB or at $\text{SSR}=10$\,dB.
For the case of $\SpfN=2$ with $\text{SSR}=0$\,dB, it is again based purely on luck whether the baseline receiver 
acquires the legitimate signals or the spoofed signals. Hence, there is a $25\%$ probability of acquiring 
the legitimate rather than the spoofed signals in both cases, as reflected by the $25\%$ success rate of positioning. 
In all other cases, the baseline exhibits only negligible chances of correct positioning. 
In contrast, \gls{schieber} is again able to reliable reject the spoofed signals and to achieve 
positioning with comparable performance as in the unobstructed case.

\subsubsection{Simultaneous Jamming and Spoofing}
\begin{figure}[tp]
    \centering
    \includegraphics[width=\figwidth]{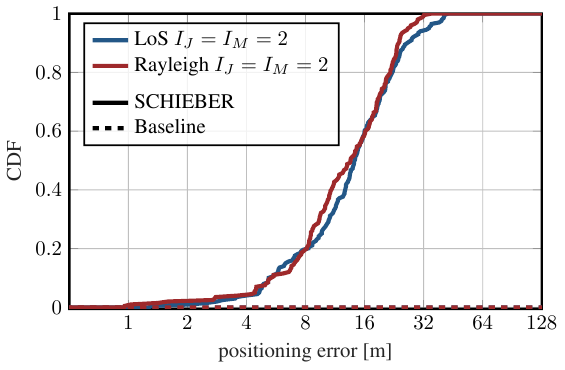}
    \caption{Cumulative distribution function (CDF) of the positioning error of SCHIEBER and the corresponding baseline in the under simultaneous jamming and spoofing.}
    \label{fig:simultaneous}
\end{figure}
Finally, we consider the performance against simultaneous jamming and spoofing. 
Moreover, in this experiment, we also consider what happens when the attackers' channels exhibit i.i.d. 
Rayleigh fading rather than LoS characteristics. 
\fref{fig:simultaneous} shows the performance (at $\text{SNR}=-20$\,dB) against $\JamN=2$ jammers 
with a \gls{jsr} of $30$\,dB and $\SpfN=2$ spoofers with an \gls{ssr} of $10$\,dB. 
The results show that in both cases, the baseline fails completely while \gls{schieber} 
is able to mitigate the jammers as well as the spoofers, and still achieves the same performance as
in the unobstructed case.

\section{Conclusions}
We have proposed \gls{schieber}, a novel multi-antenna based method for resilient GNSS positioning under both 
multi-antenna jamming and multi-antenna spoofing. 
Compared to traditional GNSS receivers, \gls{schieber} uses a modifed signal acquisition stage 
that mitigates jammers based on adaptive spatial filtering. 
\gls{schieber} then rejects spoofed signals based on a novel 
receiver-position-invariant pairwise plausibility test. 
The efficacy of \gls{schieber} is demonstrated via extensive systems 
simulation and for a range of different attack scenarios. 

The main advantages of \gls{schieber} compared to the state of the art
is that it makes minimal assumptions about the attackers (in particular, 
the number and type of attackers need not be known, no assumptions
about the receive power of the attackers' signals are made, and 
the channels of the attackers can be \gls{los} as well as Rayleigh fading)
and that it requires \emph{no} a priori estimate of the receiver position. 
Limitations to be addressed in future work are the assumption of a pure
\gls{los} link between the receiver and the legitimate satellites
as well as the limited positioning precision due to integer oversampling
and the absence of tracking and smoothing over time.


\balance

\end{document}